\documentclass[authoryear,preprint,3p,12pt]{elsarticle}
\usepackage{dcolumn}
\usepackage{bm}
\usepackage{amsfonts}
\usepackage{amsmath}
\usepackage{textcomp}
\usepackage{amssymb}
\usepackage{amsbsy}
\usepackage{graphicx}
\usepackage{natbib}
\usepackage{subfigure}
\usepackage{color}

\usepackage{mathabx}

\usepackage{graphicx}
\usepackage{mathrsfs} % Para las letras barrocas
\usepackage{lineno}
\usepackage{hyperref}

\usepackage[normalem]{ulem}
\usepackage{multirow}

\catcode`\@=11
\def\gtsim{\mathrel{\vcenter{\m@th\offinterlineskip
\hbox{$\hfill>\hfill$}\kern.5ex\hbox{$\hfill\sim\hfill$}}}}
\catcode`\@=12

\catcode`\@=11
\def\ltsim{\mathrel{\vcenter{\m@th\offinterlineskip
\hbox{$\hfill<\hfill$}\kern.5ex\hbox{$\hfill\sim\hfill$}}}}
\catcode`\@=12 \catcode`\@=12

\journal{International Journal of Multiphase Flow}

\begin{document}

\begin{frontmatter}
\title{Bubble formation regimes in forced co-axial air-water jets}
%\title{Bubble formation regimes in forced co-axial air-water jets\tnoteref{t1}}
%\tnotetext[t1]{This work has been supported by the Spanish MINECO and European Funds under projects DPI2017-88201-C3-2-R and DPI2017-88201-C3-3-R. JRR wants to acknowledge the Spanish MINECO for the financial support provided by the Fellowship BES-2015-071329.}
%\tnotetext[t2]{The second title footnote which is a longer text matter to fill through the whole text width and overflow into another line in the footnotes area of the first page.}
\author[1]{J. Ruiz-Rus\corref{cor1}}
\ead{jrrus@ujaen.es}
%\address{\'Area de Mec\'anica de Fluidos, Departamento de Ingenier\'{\i}a Mec\'anica y Minera. Universidad de Ja\'en. Campus de las Lagunillas, 23071, Ja\'en, Spain.}
\author[1]{R. Bola\~nos-Jim\'enez}
\ead{rbolanos@ujaen.es}
%\address{\'Area de Mec\'anica de Fluidos, Departamento de Ingenier\'{\i}a Mec\'anica y Minera. Universidad de Ja\'en. Campus de las Lagunillas, 23071, Ja\'en, Spain.}
\author[2]{A. Sevilla}
\ead{asevilla@ing.uc3m.es}
\author[1]{C. Mart\'inez-Baz\'an}
\ead{cmbazan@ujaen.es}
\address[1]{\'Area de Mec\'anica de Fluidos, Departamento de Ingenier\'{\i}a Mec\'anica y Minera. Universidad de Ja\'en. Campus de las Lagunillas, 23071, Ja\'en, Spain.}
\address[2]{\'Area de Mec\'anica de Fluidos, Departamento de Ingenier\'{\i}a T\'ermica y de Fluidos. Universidad Carlos III de Madrid, 28911, Legan\'es, Spain.}

\cortext[cor1]{Corresponding author}

\begin{abstract}
We report a detailed experimental characterization of the periodic bubbling regimes that take place in an axisymmetric air-water jet when the inner air stream is forced by periodic modulations of the pressure at the upstream air feeding chamber. When the forcing pressure amplitude is larger than a certain critical value, the bubble formation process is effectively driven by the selected frequency, leading to the formation of nearly monodisperse bubbles whose volume is reduced by increasing the forcing frequency. We reveal the existence of two different breakup modes, M1 and M2, under effective forcing conditions. The bubble formation in mode M1 resembles the natural bubbling process, featuring an initial radial expansion of an air ligament attached to the injector, whose initial length is smaller than the wavelength of a small interfacial perturbation induced by the oscillating air flow rate. The expansion stage is followed by a ligament collapse stage, which begins with the formation of an incipient neck that propagates downstream while collapsing radially inwards, leading to the pinch-off of a new bubble. These two stages take place faster than in the unforced case as a consequence of the the air flow modulation induced by the forcing system. The breakup mode M2 takes place with an intact ligament longer than one disturbance wavelength, whereby the interface already presents a local necking region at pinch-off, and leads to the formation of bubbles from the tip of an elongated air filament without an expansion stage. Scaling laws that provide closed expressions for the bubble volume, the intact ligament length, and the transition from the M1 breakup mode to the M2, as functions of the relevant governing parameters, are deduced from the experimental data. In particular, it has been found that the transition from mode M1 to mode M2 occurs at $(St_f \Lambda W\!e)_c=0.25$  and that the intact ligament scales as $l_i/r_o \sim St_f^{-1} \Lambda^{1/5} W\!e^{1/4}$ within the breakup mode M1. Here $r_o$ is the radius of the gas stream, $\Lambda$ the water-to-air velocity ratio, $We$ the Weber number and $St_f$ the dimensionless forcing frequency.

\end{abstract}
\end{frontmatter}
%\linenumbers
%--------------------------------------------------------------------------------%
%----------------------------------INTRODUCTION----------------------------------%
%--------------------------------------------------------------------------------%
\section{Introduction}\label{COFLOW2019_section:introduction}

Bubble generation represents an important operation in the context of material, mineral, chemical and food industries, to name a few. Many emerging technologies, such as those related to biomedicine, require the use of monodisperse microbubbles~\citep{RodriguezARFM2015}. In this context, bubbles can be used, for instance, as contrast agents in ultrasound techniques or for drug delivery~\citep{Ferrara_2007}. These emerging applications are demanding new methods to generate bubbles at a given frequency and with a highly controllable size. Water aeration is another important application, characterized by the requirement of very large air throughputs~\citep[see][]{Amand_Review_2013}. The efficiency of these systems is governed by the amount of gas dissolved in the liquid per unit time, and the dissolution rate is proportional to the surface of the gas-liquid interface. A possible way to increase the interfacial area is to decrease the volume of the injected bubbles, what can in principle be achieved using porous plates or microfluidic devices. However, these systems are prone to clogging issues due to the small size of the injection ports. Thus, the development of robust devices, able to inject large air flow rates with small bubbles is still a challenging engineering problem.

The simplest method to control the bubble size and formation frequency is the parallel co-flow configuration~\citep[see][among others]{Chuang1970,OguzJFM93,Tsuge_1997,Bhunia_1998,SevillaJFM2005,Gordillo_2007}, where the injected gas flow discharges inside a co-flowing laminar liquid stream. Compared with the case of injecting air in still liquids, the co-flow technique enables the generation of smaller and monodisperse bubbles, allowing the injection of higher gas flow rates, and preventing the bubble coalescence at the exit of the injector. Both the classical cylindrical~\citep[see][and references therein]{RodriguezARFM2015} and the alternative planar~\citep{Gutierrez-MontesIJMF2013, Bolanos_IJMF2016,Ruiz-Rus_2017} geometries have been studied, and the influence of the gas injection conditions on the bubbling processes has been described~\citep{Gordillo_2007,Gutierrez-MontesIJMF2014}. In particular,~\citet{SevillaJFM2005} and later on~\citet{Gordillo_2007}, performed detailed experimental, theoretical and numerical studies of the formation of bubbles in a cylindrical air-water co-flowing jet at high Reynolds numbers and provided suitable scaling laws for the bubble size and the bubbling time. In fact, under these conditions, the mechanism driving the bubble formation is purely inertial, and the production frequency of the emitted bubbles is roughly proportional to the inverse of liquid convective time, $f_b \, \propto \, u_w/r_o$, where $u_w$ is the velocity of the outer co-flowing liquid stream and $r_o$ is the gas injector radius~\citep{RodriguezARFM2015}. Thus, generating very small bubbles at very large frequencies, a typical requirement of many technological applications, implies the use of small injectors and large liquid co-flow velocities, what makes the system energetically costly. In addition, in both the cases of constant gas flow rate and constant injection pressure, a dependence on the gas flow rate has been found~\citep{RodriguezARFM2015}, which hinders the independent control of the bubble volume and formation frequency.

A natural way to control the formation of bubbles at the frequencies and sizes required for practical applications is to apply an appropriate forcing protocol to the gas-liquid stream. Such techniques have been widely studied and successfully applied to the case of liquid jets injected into ambient gas, mostly motivated by the practical need to improve the design of inkjet printing devices~\citep{Lee_JResDev_1974,Basaran2002}. However, it is important to emphasize that, to generate a slender liquid jet inside a gas, the only condition that must be accomplished is that the inertia of the injected liquid stream is greater than the surface tension forces that tend to hold the liquid mass attached to the injector walls. In dimensionless terms, this condition indicates that the liquid Weber number must be of order unity or larger, in which case the slender liquid jet is convectively unstable~\citep{LeibyGoldsteinAC,LeibyGoldstein}. In contrast, when a gas stream is injected subsonically into an ambient liquid at rest, the resulting flow regime is absolutely unstable, since the inertia of the inner gas is always much smaller than that of the outer liquid. As a result, in this \emph{bubbling regime}, large individual bubbles are formed near the injector tip~\citep{OguzJFM93}. However, when an outer liquid co-flow is applied with a sufficiently large velocity, a transition to a {jetting regime} takes place, featuring a long gas jet that breaks-up into bubbles far from the nozzle~\citep{Sevilla_2005_PoF}. From a practical point of view, the main disadvantage of the jetting regime is that it requires a large liquid velocity, of the order of the gas velocity, and therefore it finds little use in applications. The fact that the flow is dominated by the inertia of the outer liquid has another important consequence for the physics of bubble generation. Indeed, the typical pressure fluctuations inside a forming bubble that is attached to the injector scale with the outer liquid density. These pressure fluctuations may well be of the order of the pressure drop along the gas feeding line, which scales with the gas density, the gas viscosity, and the geometry of the injection line. In these common cases, the injected gas flow rate may vary with time due to the forcing of the pressure disturbances at the injector outlet~\citep{Gordillo_2007}. Moreover, for modeling purposes, the unsteady gas flow in the feeding line is unknown \emph{a priori}, and must be included as an additional unknown together with a proper model to account for the hydraulic resistance associated with the gas feeding system~\citep{OguzJFM93,Gordillo_2007}. From these considerations, it is clear that the design of forcing strategies for the controlled formation of bubbles is more challenging than in the case of liquid jets.

Thus, due to the relevance of the topic, during the last decades a considerable effort has been devoted to develop new methods to generate controlled-size bubbles using different forcing techniques. The application of pulsed pressure waves to the gas phase~\citep{Shirota_PoF_2008,Najafi_2008,Makuta_Ultrasonic_2013}, the movement or vibration of the gas injector~\citep{Grinis_1999,Vejrazka_FDR_2008,Waghmare_IndEngChemRes_2008,Wang_2016}, the use of a rotational porous plate~\citep{Fujikawa_2003}, elastic tubes~\citep{Sanada_2013,Abe_ChemEngSci_2015} or a piston~\citep{Ostmann_2018}, or the application of electric fields~\citep{DiBari_2013,Xu_2017,Wang_2018} constitute some of the methods used to control the bubble size when the air discharges into still liquid. 

However, to the best of our knowledge, a forcing technique that stimulates the inner air stream in a cylindrical co-flow configuration has not been explored yet. In a previous related work,~\citet{Ruiz-Rus_2017} demonstrated that the modulation of the external liquid flow rate in a planar co-flow configuration allows to effectively and independently control both the generation frequency and the bubble size. In the present work, the forcing is applied to the inner gas stream in a cylindrical co-flow configuration. 

Consequently, the purpose of the present study is to assess the performance of a new method to effectively and easily control both the size and frequency of the bubble generation process in a cylindrical co-flow configuration. To that end, we experimentally analyze the mechanisms governing the bubble formation dynamics under the influence of harmonic pressure perturbations induced in the gas feeding line. The paper is organized as follows. Section~\ref{COFLOW2019_section:Exp_App} describes the experimental facility and techniques; Section~\ref{COFLOW2019_section:Results} reports the results of the bubbling process obtained by imposing the effective forcing amplitude, featuring the presence of two different bubbling modes. Section~\ref{COFLOW2019_section:Char_Length} focuses on the characterization of the bubble intact length observed in each bubbling mode, as well as on the transition between bubbling modes. Finally, Section~\ref{COFLOW2019_section:Conclusions} is devoted to conclusions.

%--------------------------------------------------------------------------------%
%---------------------------------EXPERIMENTS----------------------------------%
%--------------------------------------------------------------------------------%
\section{Experimental approach}\label{COFLOW2019_section:Exp_App}

In this section, the experimental setup, the flow conditions and the experimental techniques used to obtain the results are described in detail.

\subsection{Facility and flow conditions}\label{COFLOW2019_subsection:Exp_Setup}
\begin{figure}[t]
	\centering
    \includegraphics[width=1\textwidth]{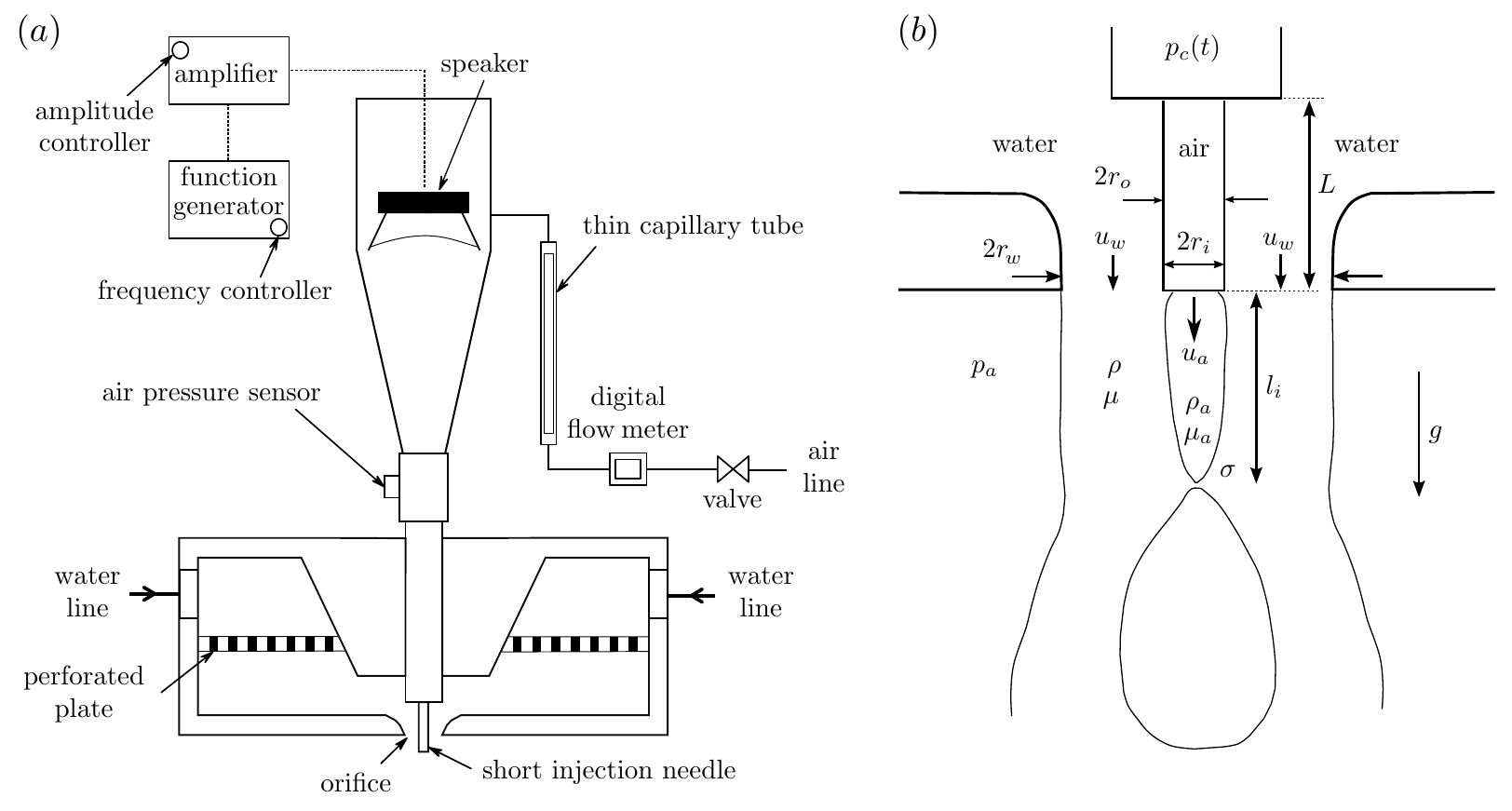}
    \caption{($a$) Schematic representation of the experimental facility, showing the feeding systems of both the water and the air streams. ($b$) Sketch of the analyzed flow configuration, including the relevant physical parameters. The bubble pinch-off instant is depicted, showing the intact ligament attached to the needle, as well as the newly formed bubble.}
    \label{COFLOW2019_fig:Scheme}
\end{figure}

The experimental set-up used in the present work, schematized in Fig.~\ref{COFLOW2019_fig:Scheme}($a$), consists of a gas-liquid co-flow device which incorporates a forcing system to stimulate the air stream. In this facility, a gas stream is coaxially injected inside a free liquid jet at high Reynolds and Weber numbers. The continuous water jet discharges vertically through an orifice of radius $r_w = 4$ mm from a liquid reservoir. A constant water flow rate, $Q_w$, controlled and measured by a high-precision valve and a flow meter respectively, was supplied to the reservoir through a water line. To attenuate undesired liquid disturbances, a perforated plate and a piece of foam were placed inside the water vessel. Moreover, the orifice was shaped to ensure a nearly uniform water velocity profile at the exit. The air flow was injected at the centerline of the water jet through a needle with inner and outer radii $r_i = 0.4$ and $r_o = 0.6$ mm, respectively. Besides, a short needle length, $L = 17$ mm, was used to minimize the associated pressure losses and the damping of the pressure perturbations introduced upstream of the needle by the forcing system. It is important to emphasize that, with this set-up, the air flow rate feeding the bubble does not remain constant during the bubble formation process. In fact, the associated pressure drop along the needle is lower than the characteristic pressure fluctuations inside the forming bubble, causing the instantaneous air flow rate feeding the bubble, $Q_a(t)$, to vary during the formation cycle, even in the unperturbed bubbling cases~\citep{Gordillo_2007}.

The co-flow system described above also incorporated a forcing device into the air feeding line, with the aim at inducing a controlled pressure modulation to the air stream (Fig.~\ref{COFLOW2019_fig:Scheme}$a$). The forcing system consists of a loudspeaker (SP-45$/$8, MONACOR) placed inside a funnel-shaped vessel that includes an air inlet and a needle holder at its outlet. The inner shape of the air chamber allows an optimal performance of the speaker, minimizing the attenuation of the induced pressure perturbations. Contrary to the concept proposed by~\citet{Shirota_PoF_2008}, the present device is continuously fed with a constant air flow rate, $Q_c$. Therefore, in the work at hand, a train of bubbles of controlled volume could be generated at the forcing frequency, rather than a single bubble produced by a pair of pulsed acoustic waves~\citep{Shirota_PoF_2008, Abe_ChemEngSci_2015, RodriguezARFM2015}. The air flow was supplied to the chamber from a compressed air bottle, controlled with a pressure regulator and a high-resolution valve, and measured with a digital mass flow meter. Moreover, a long capillary tube was placed upstream of the chamber, imposing a pressure drop much larger than the characteristic pressure variation inside the chamber (see~\citealt{Corchero_2006}), thus ensuring that air was supplied to the chamber at a constant flow rate, $Q_c$. Finally, the speaker was excited by means of periodic sinusoidal electric signals provided by a tunable-frequency function generator (FG110, YOKOGAWA), which allowed an easy selection of the forcing frequency, $f_{\!f}$. These signals, previously amplified by a power amplifier (PA-702, MONACOR), drive the speaker, whose motion induces a periodic perturbation of the air pressure inside the chamber, $p_c(t)$. To fully characterize the induced perturbation, $p_c(t)$ was measured by a pressure sensor placed at the chamber exit (Fig.~\ref{COFLOW2019_fig:Scheme}$a$). The pressure signals were registered during the speaker performance, showing sinusoidal shapes with the forcing frequency, $f_{\!f}$, and amplitude, $\Delta p_c$, accurately adjustable through the amplifier (see Fig.~\ref{COFLOW2019_fig:Measures}).

Fig.~\ref{COFLOW2019_fig:Scheme}($b$) shows a schematic representation of the bubble pinch-off instant, including the relevant physical parameters involved in the problem, where $p_a$ denotes the atmospheric pressure, $g$ is the acceleration of gravity, $\sigma$ is the air-water surface tension coefficient, $\mu$ and $\mu_a$ are the water and air viscosities, respectively, $\rho$ represents the water density while $\rho_a$ stands for the air density, and $u_w$ and $u_a$ are the mean water and air velocities at the exit, respectively. Regarding the geometrical parameters, $r_i$ and $r_o$ stand for the inner and the outer air injector radii, $L$ is its length, and $r_w$ is the water jet radius. Besides, $l_i$ indicates the intact length, which is the length of the air lump that remains attached to the needle tip after the bubble detachment. In this work, the water jet velocity was varied in the range $1 < u_w=Q_w/A_w < 2$ m/s, being $A_w=\pi(r_w^2-r_o^2)$ the water exit cross-section. Although the instantaneous air flow rate $Q_a(t) \neq Q_c$ in all the experiments reported here, its time average over one bubbling period $T$ accomplishes $1/T\int_0^T Q_a(t)\,{\rm d}t=Q_c$. Thus, the average air velocity at the needle exit is defined as $u_a = Q_c / (\pi r_o^2)$, and was modified in the range $0.5 < u_a < 8$ m/s. In addition, the forcing frequency was varied in the range of $250 \leq f_{\!f} \leq 500$ Hz, being the lower limit imposed by the natural bubbling frequency, $f_n$, while the upper value depends on the maximum response of the loudspeaker at the selected frequency. Using $r_o$ and $u_w$ as the characteristic scales for length and velocity, respectively, the dimensionless parameters describing the problem are the following. The liquid Reynolds and Weber numbers, $Re = \rho u_w r_o / \mu \gg 1$ and $W\!e = \rho u_w^2 r_o / \sigma \gg 1$ which, as mentioned before, are both very large indicating that the inertia of the liquid dominates over viscous and surface tension forces in the experiments reported herein. The Froude number based on the intact length, $F\!r = g l_i / u_w^2 \ll 1$, inferring that gravity effects are negligible in all the experiments. The value of the air Reynolds number, $Re_a = \rho_a u_a r_o / \mu_a$, was always sufficiently large for viscous effects to be negligible within the bubble. In addition, the air flow inside the needle was not fully developed~\citep{Gordillo_2007}. Therefore, the dimensionless control parameters of the problem are reduced to the forcing Strouhal number, $St_{\!f} = f_{\!f} r_o / u_w$, and the liquid-to-gas mean velocity ratio, $\Lambda = u_w / u_a$. The experimental range of both parameters covered in the present work was $0.05\lesssim St_{\!f} \lesssim 0.23$ and $0 < \Lambda< 2.3$. At this point, it is important to mention that the unperturbed experiments reported here lead to well defined natural bubbling regimes with periodic generation of monodisperse bubbles~\citep{SevillaJFM2005,Gordillo_2007}. In fact, these flow conditions give rise to natural bubble generation frequencies in the range $125 < f_n < 300$ Hz. However, larger values of $\Lambda$ lead to the transition from periodic bubbling to jetting, as reported by~\citet{Sevilla_2005_PoF} for constant air flow rate conditions. 

\subsection{Experimental methods}\label{COFLOW2019_subsection:Exp_Meth}
\begin{figure}%[t]
	\centering
    \includegraphics[width=0.8\textwidth]{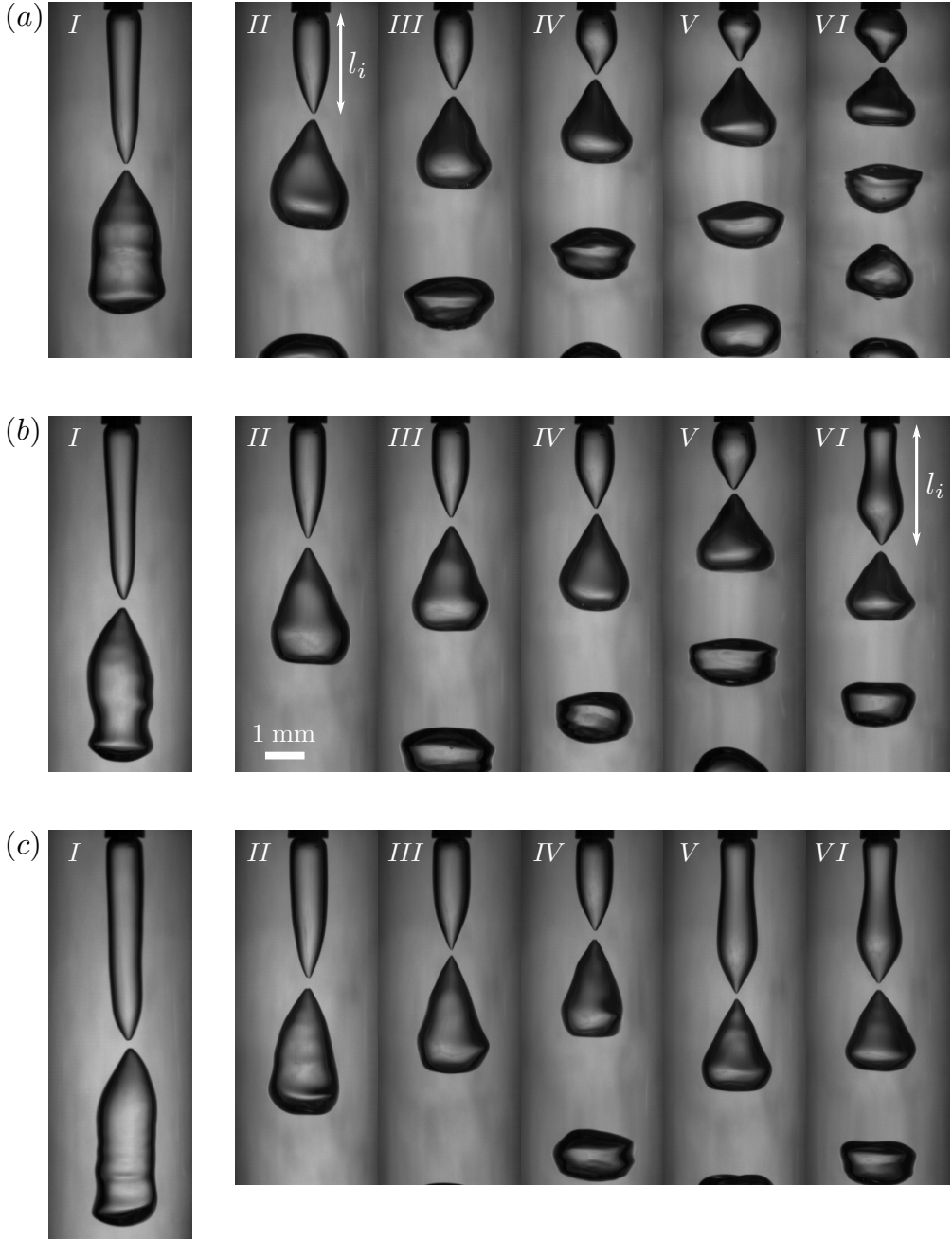}
    \caption{Snapshots of different bubbling regimes just after bubble pinch-off. The selected flow conditions, $u_a = 2.22$ m$/$s and ($a$) $u_w = 1.53$ m$/$s ($W\!e= 19.5$, $\Lambda= 0.69$), ($b$) $u_w = 1.70$ m$/$s ($W\!e= 24$, $\Lambda= 0.76$) and ($c$) $u_w = 1.86$ m$/$s ($W\!e= 28.83$, $\Lambda= 0.84$), give rise to the natural bubbling frequencies $f_n = 182$; $186$ and $195$ Hz, respectively, corresponding to the unperturbed case represented in the first image of each row ($I$). The rest of the images show forced cases of the corresponding natural regime, ($I\!I$) $f_{\!f} = 250$ Hz ($St_{\!f}$= 0.098, 0.088 and 0.081 in ($a$), ($b$) and ($c$) respectively), ($I\!I\!I$) $f_{\!f} = 300$ Hz ($St_{\!f}$= 0.117, 0.106 and 0.097 in ($a$), ($b$) and ($c$) respectively), ($IV$) $f_{\!f} = 350$ Hz ($St_{\!f}$= 0.137, 0.123 and 0.113 in ($a$), ($b$) and ($c$) respectively), ($V$) $f_{\!f} = 400$ Hz ($St_{\!f}$= 0.157, 0.141 and 0.129 in ($a$), ($b$) and ($c$) respectively) and ($V\!I$) $f_{\!f} = 450$ Hz ($St_{\!f}$= 0.176, 0.159 and 0.145 in ($a$), ($b$) and ($c$) respectively). The scale bar is $1$ mm long, indicating the spatial resolution of all the shown snapshots.}
    \label{COFLOW2019_fig:Images}
\end{figure}
The experiments were conducted following a procedure similar to that described in~\citet{Ruiz-Rus_2017}. Firstly, the unperturbed air-water jet was generated by setting both the mean air and water flow velocities, $u_a$ and $u_w$. This gives rise to a natural bubbling regime in which bubbles generate at a certain frequency, $f_n$, from the pinch-off position, i.e.~the intact length, $l_i$ (see snapshots $I$ in Fig.~\ref{COFLOW2019_fig:Images}). Then, each natural case was forced by imposing a pressure modulation at a desired forcing frequency, larger than the natural one, $f_{\!f} > f_n$. For each value of $f_{\!f}$, the forcing amplitude was slowly and monotonically increased until a synchronized and periodic bubble generation at the forcing frequency was achieved (see snapshots $I\!I$-$V\!I$ in Fig.~\ref{COFLOW2019_fig:Images}), attending to the criteria established in Sect.~\ref{COFLOW2019_subsection:Exp_Criteria}. The same procedure was followed for increasing values of $f_{\!f}$ and all the natural cases investigated. Note from images in Fig.~\ref{COFLOW2019_fig:Images} that the forcing mechanism, when effective, allows us to generate bubbles at frequencies higher than the natural one, reducing thus the bubble size. Finally, measurements were performed from high-speed visualizations acquired with a backlighting technique. To that aim, images of the whole bubble interface were recorded with a Photron high-speed camera at frame rates from $37\,500$ f.p.s.~with a resolution of $192 \times 576$ pixels to $45\,000$ f.p.s.~with a $192 \times 480$ pixel resolution. A shutter speed of 8.5 $\mu$s was sufficiently fast to avoid blurred profiles even at the extremely fast bubble pinch-off event. Spatial resolutions between 23 and 40 $\mu$m/pixel were achieved using a Sigma 105 mm macro lens, depending on the size of the field of view used to simultaneously capture both the entire intact air ligament and the whole detached bubble silhouette at the pinch-off instant. In addition to the recorded high-speed images, the instantaneous air pressure inside the chamber, $p_c(t)$, was registered with an acquisition rate ten times larger than the image recording frame rate. The images and the pressure signal were synchronized through a data logger system.

Each registered image of the forming bubble was digitally analyzed using an in-house two-step interface detection routine, similar to that described by~\citet{Hijano_2015}. Firstly, a rough detection of the edges of the needle tip and of the forming bubble is performed at the pixel level. Given the non-uniformity of the background gray level (see e.g.~Fig.~\ref{COFLOW2019_fig:Images}), that makes inadequate the global binarization-threshold methods~\citep{Gonzalez_2002}, an edge detection method based on the local gray level gradient was specifically developed. In this method, the location of the pixels corresponding to the bubble interface was obtained by means of the Canny algorithm~\citep{Canny_1986}. Secondly, the accuracy of the detected contour was improved to the sub-pixel level. To that aim, a local threshold criterion was implemented by fitting a sigmoid function to the gray intensity level along the axis normal to the edge at each point of the interface~\citep{Song_1996}. More detailed information about the local threshold selection and the smoothing of the obtained interface can be found in~\cite{Vega_2009}.

\begin{figure}[t]
	\centering
    \includegraphics[width=0.7\textwidth]{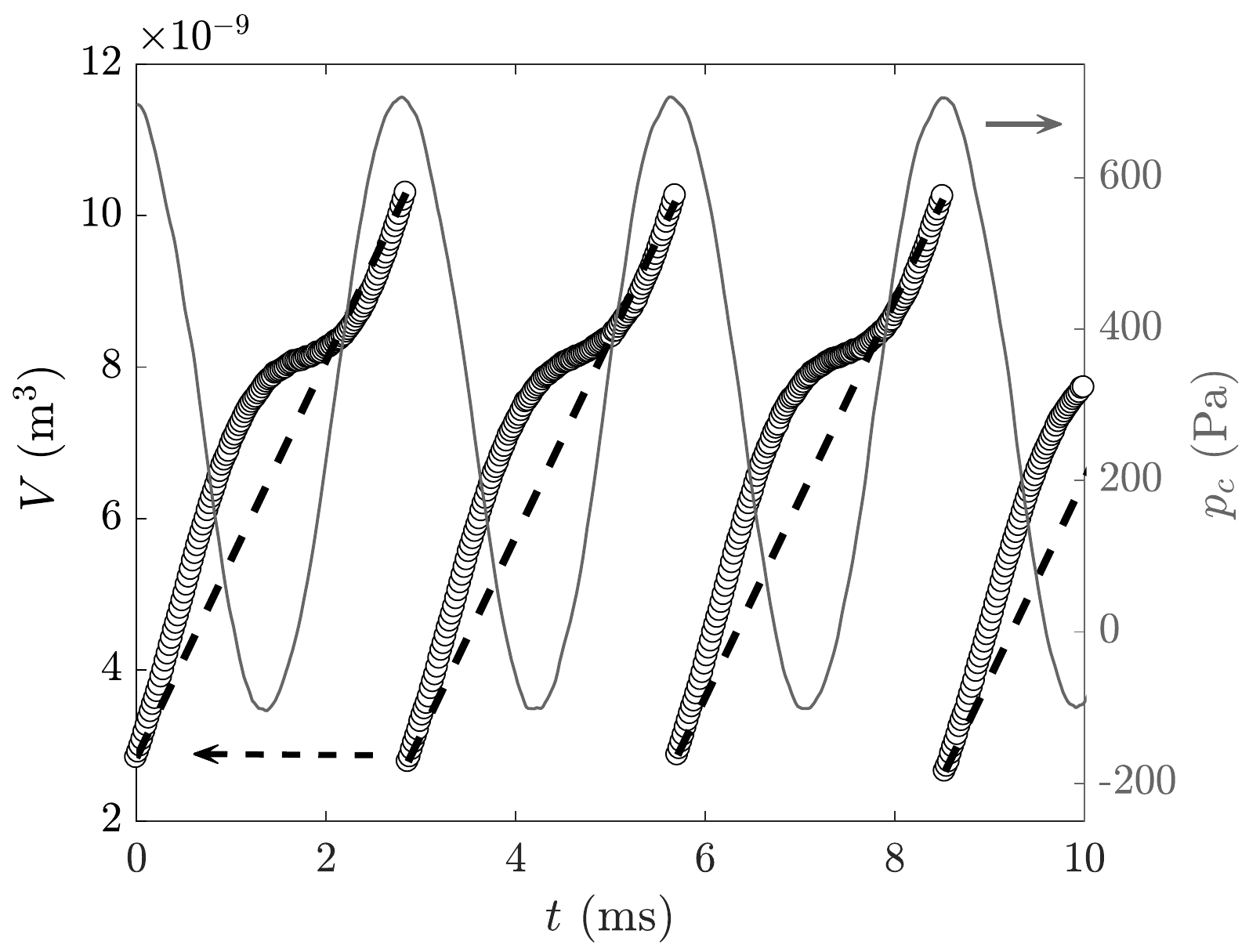}
    \caption{Time evolution of the forming bubble volume and the air pressure inside the chamber during several bubbling cycles for $u_a = 2.22$ m$/$s and $u_w = 1.70$ m$/$s ($W\!e= 24$, $\Lambda= 0.76$), forced at $f_{\!f} = 350$ Hz ($St_{\!f}= 0.123$), corresponding to panel $IV$ in Fig.~\ref{COFLOW2019_fig:Images}($b$). Symbols represent the instantaneous volume of the forming bubble extracted from the image analysis (left axis). The dashed lines indicate the time evolution of the bubble volume that would take place under constant air flow rate conditions. The solid line displays the pressure registered at the feeding chamber, showing a nearly sinusoidal shape at the selected forcing frequency, $f_{\!f}$ (right axis).}
    \label{COFLOW2019_fig:Measures}
\end{figure}
The bubble contour detected in each frame allowed us to obtain the instantaneous volume of the growing air cavity attached to the injector, under the assumption of axisymmetry, as $V(t)= \int_{0}^{z_t}\pi D^2(z,t)/4\;{\rm d}z$. Here, $D(z,t)$ denotes the diameter of the interface obtained from the image, $z$ is the axial coordinate measured from the injector tip, and $z_t$ represents the bubble tip position, defined as the downstream location at which the interface diameter becomes zero. Figure~\ref{COFLOW2019_fig:Measures} represents the time evolution of the growing cavity volume obtained from the images (left axis), together with the synchronized chamber pressure signal (right axis), for the experimental flow conditions $u_a = 2.22$ m$/$s and $u_w = 1.70$ m$/$s ($W\!e= 24$, $\Lambda= 0.76$), forced at $f_{\!f} = 350$ Hz ($St_{\!f}= 0.123$). The symbols in Fig.~\ref{COFLOW2019_fig:Measures} show the experimental volume growing from the initial air lump that remains attached to the injector once the previous bubble pinches-off, referred to as the \emph{intact ligament}. It can be observed that, as an effect of the forcing process, the volume evolution during the bubbling cycle is quite different from that which would take place under constant air flow rate conditions, depicted by dashed lines in Fig.~\ref{COFLOW2019_fig:Measures}. At the end of the bubbling cycle, the volume of the air cavity attached to the injector abruptly decreases to reach the volume of the intact ligament, indicating the detachment of a bubble and the beginning of a new cycle. The axial position at which the pinch-off occurs is the \emph{intact length}, $l_i$, defined as the value of $z_t$ at the first frame of each cycle. Under periodic bubbling conditions, the time elapsed between two consecutive bubble detachments is just the inverse of the bubbling frequency $f_b$. In addition, the volume of the generated bubble, $V_b$, is obtained as the difference between the volume measured at the last frame of the cycle and that of the corresponding intact ligament. Moreover, the amplitude of the induced chamber pressure, $\Delta p_c$, has been obtained by computing the mean value of the pressure signal envelope through a Hilbert transform, as described in~\citet{Jimenez-Gonzalez_2017}. Finally, the averaged values of $l_i$, $f_b$ and $V_b$, as well as their standard deviations, were obtained by analysing a minimum of 20 bubbling events for each selected value of the forcing amplitude $\Delta p_c$.

\subsection{Experimental procedure: criteria to determine the effective forcing amplitude}\label{COFLOW2019_subsection:Exp_Criteria}

\begin{figure}[t]
	\centering
    \includegraphics[width=0.85\textwidth]{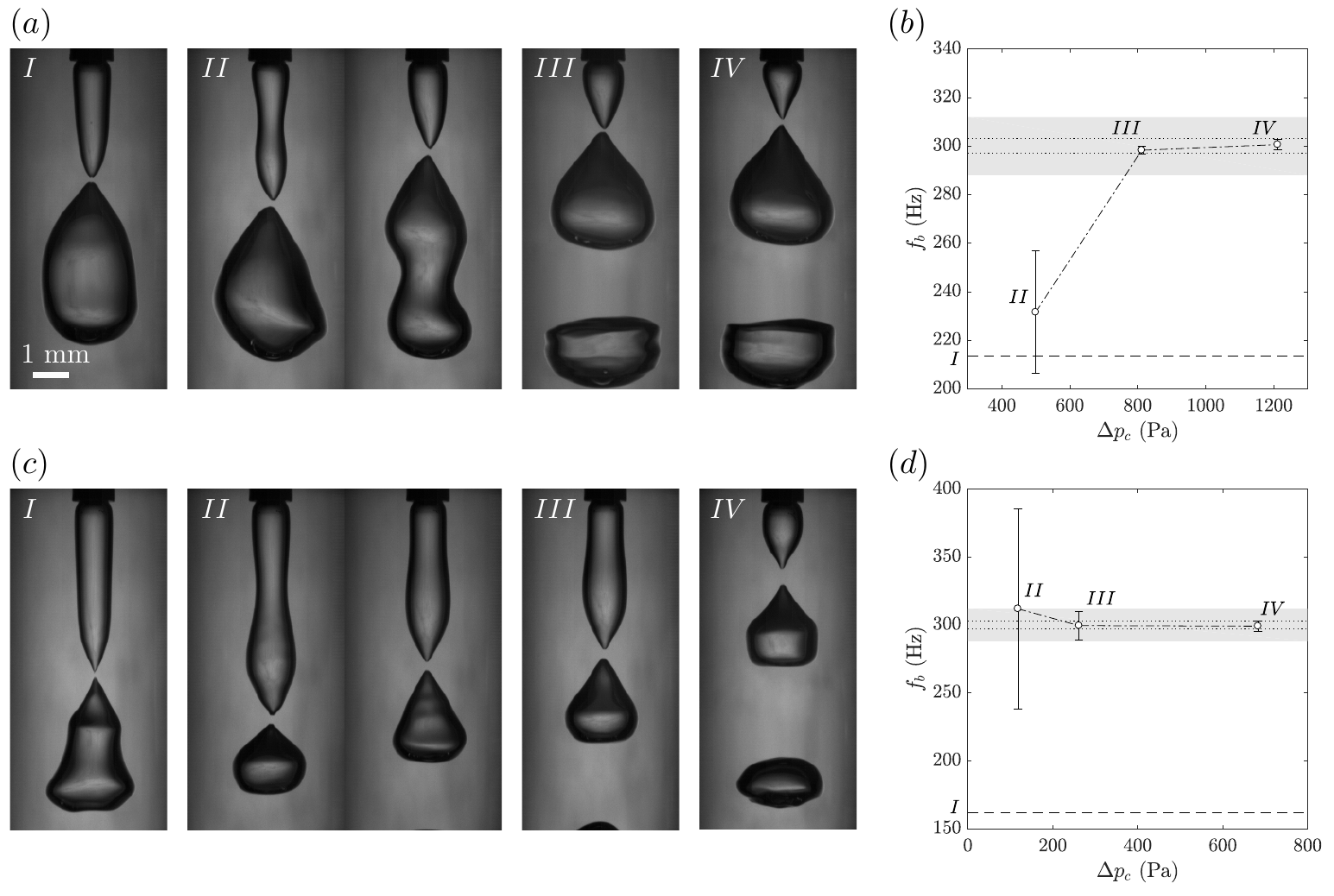}
    \caption{($a$) Experimental images of the instant just after the bubble pinch-off for the flow conditions $u_a = 3.80$ m$/$s and $u_w = 1.36$ m$/$s ($W\!e= 15.41$, $\Lambda= 0.358$) corresponding to the natural case, $f_n = 213$ Hz ($I$), and forced cases with increasing pressure amplitude at 300 Hz ($I\!I$-$IV$) ($St_{\!f}$= 0.132). ($b$) Mean bubbling frequency of forced cases in ($a$) as a function of the pressure amplitude, whose associated standard deviation is plotted with error bars. The natural bubbling frequency is indicated with a horizontal dashed line. The range of admissible values of the mean bubbling frequency and its standard deviation, attending to the effectiveness criteria, are depicted as horizontal dotted lines and as a shaded region, respectively. Both snapshots in ($I\!I$) correspond to two consecutive bubble detachments of an experiment that is forced with a non-effective amplitude. Effective forcing processes under the present conditions exhibit the so-called breakup mode M1, being the corresponding critical amplitude represented in ($I\!I\!I$), while a larger value is shown in ($IV$). ($c,d$) Same as ($a,b$) but for a case with $u_a = u_w = 1.36$ m$/$s ($W\!e= 15.41$, $\Lambda= 1$), $f_n = 162$ Hz. Under these conditions ($St_{\!f}$= 0.132), the critical pressure amplitude leads to an effective forcing process which responds in the breakup mode M2 ($I\!I\!I$). For amplitudes larger than the critical one, although effective, the forcing process departs from M2 ($IV$).}
    \label{COFLOW2019_fig:Effective}
\end{figure}
The experimental procedure described in the previous section implies that only the forced cases leading to synchronized bubble generation at the forcing frequency, are considered in the present work. These cases, denoted as \emph{effective} from now on, are achieved if the forcing amplitude is sufficiently large, a value that depends on the unperturbed flow conditions as well as on the forcing frequency. The minimum pressure amplitude required to achieve an effective forcing process will hereinafter be called the \emph{critical amplitude}. For amplitudes smaller than the critical one, the forcing is not able to properly control the bubble generation process. Therefore, to clearly distinguish between both situations, the bubbling regimes established for each selected value of $\Delta p_c$, were analyzed following an effectiveness criterion based on the measured bubbling frequency, as described in detail below. At this point, it should be mentioned that, depending on the flow conditions and on the forcing frequency, two different forced bubbling modes were found, denoted as breakup modes M1 and M2, respectively (see e.g.~snapshots ($a$)$I\!I$-$V\!I$; ($b$)$I\!I$-$V$ and ($c$)$I\!I$-$IV$ for M1, as well as ($b$)$V\!I$ and ($c$)$V$-$V\!I$ for M2, in Fig.~\ref{COFLOW2019_fig:Images}). Briefly stated, mode M1 is characterized by the periodic formation of bubbles at a distance from the injection needle smaller than $\lambda_{\!f}= u_w/f_{\!f}$, which can be interpreted as the wavelength of a small-amplitude interfacial disturbance travelling downstream at the liquid velocity. In contrast, mode M2 is characterized by the formation of bubbles at a distance larger than $\lambda_{\!f}$, featuring more than one local maximum on the interface of the ligament at the pinch-off instant. Both modes are effective for amplitudes larger than the corresponding critical ones, giving rise to the emission of monodisperse bubbles at the selected forcing frequency. However, their characteristic features, as well as the way in which the bubble pinch-off takes place, differ substantially. A detailed description of the dynamics involved in each mode is given in Sect.~\ref{COFLOW2019_subsection:Res_Bubbling}.

Figure~\ref{COFLOW2019_fig:Effective} illustrates the procedure followed to find the critical amplitude. In this case the effectiveness criterion explained in the previous paragraph was applied to different forcing processes with equal forcing frequency, $f_{\!f} = 300$ Hz, and increasing pressure amplitude. Two different flow conditions, $u_a = 3.80$ m$/$s, $u_w = 1.36$ m$/$s (Fig.~\ref{COFLOW2019_fig:Effective}$a,b$) and $u_a = u_w = 1.36$ m$/$s (Fig.~\ref{COFLOW2019_fig:Effective}$c,d$), are discussed. For amplitudes above the critical one, the former case generates bubbles under mode M1 (see snapshot $I\!I\!I$ in Fig.~\ref{COFLOW2019_fig:Effective}$a$) while, in the latter one, the bubbling regime is M2 (see snapshot $I\!I\!I$ in Fig.~\ref{COFLOW2019_fig:Effective}$b$). As anticipated above, for a forcing condition to be considered effective, the values of the mean bubbling frequency and its standard deviation obtained for each selected pressure amplitude, should be within the limits established by the effectiveness criterion. More precisely, a variation of $\pm1$\% around the target forcing frequency is accepted for the mean bubbling frequency, while a standard deviation less than 4\% results acceptable. It is worth pointing out that the ranges of admissible values established for the criteria have been selected attending to the characteristic fluctuations observed under natural bubbling conditions. Figure~\ref{COFLOW2019_fig:Effective}($b,d$) represents with symbols the corresponding mean bubbling frequency obtained under the imposed pressure amplitude, being the standard deviation plotted as error bars. The initial unperturbed bubbling frequency is indicated with a dashed line, corresponding to the natural case (see snapshots $I$ in Fig.~\ref{COFLOW2019_fig:Effective}). 

Specifically, the critical pressure amplitudes for all the experimental flow conditions, forced at different frequencies, were found following the steps described below for the conditions presented in Fig.~\ref{COFLOW2019_fig:Effective}($a,b$). We started from the flow conditions $u_a = 3.80$ m$/$s and $u_w = 1.36$ m$/$s, establishing the unperturbed bubbling regime (Fig.~\ref{COFLOW2019_fig:Effective}$a \, I$), which leads to a natural bubbling frequency of $f_n = 213$ Hz. Then, we set the desired forcing frequency, $f_{\!f} = 300$ Hz in this case, and the critical forcing amplitude was searched by increasing the pressure amplitude in small steps. If the forcing amplitude is not sufficiently large, the imposed pressure fluctuations lead to an ineffective forcing in which the bubbling process is not able to achieve the forcing frequency, producing polydisperse bubbles, as can be seen in Fig.~\ref{COFLOW2019_fig:Effective}($a \, I\!I$). This ineffective amplitude, $\Delta p_c = 500$ Pa, induces a bubbling frequency whose mean value is outside the ranges established by the effectiveness criteria (point $I\!I$ in Fig.~\ref{COFLOW2019_fig:Effective}$b$). Notice that a high standard deviation can be observed accordingly. If the amplitude is slowly increased, it reaches a certain value, $\Delta p_c = 812$ Pa in the case at hand, for which both the mean bubbling frequency and its associated standard deviation satisfy the effectiveness criteria, leading to the periodic emission of monodisperse bubbles at the forcing frequency (Fig.~\ref{COFLOW2019_fig:Effective}$a \, I\!I\!I$). This pressure amplitude corresponds to the critical one, and the associated breakup mode can be identified from the images, according to the intact ligament length. If the amplitude is further increased (Fig.~\ref{COFLOW2019_fig:Effective}$a \, IV$), the forcing mechanism remains effective, presenting a forced bubbling regime similar to that obtained at the critical amplitude, but generating bubbles closer to the needle exit due to a reduced intact length.

The steps described in the previous paragraph were followed for all the forced cases, including the conditions leading to forced bubbling regimes under the M2 breakup mode (see Fig.~\ref{COFLOW2019_fig:Effective}$c,d$). However, as explained in detail below, the M2 regime is characterized by slender intact ligaments, similar to those obtained in the natural cases with high values of $\Lambda$. This fact turns the bubbling process more susceptible to noise disturbances in the liquid stream~\citep{Sevilla_2005_PoF}, leading to weak fluctuations of the bubbling frequency, as confirmed by the slightly larger standard deviation obtained for the point ($I\!I\!I$) in Fig.~\ref{COFLOW2019_fig:Effective}($d$). However, for amplitudes larger than the critical one, although still effective, the forcing process becomes more energetic and the fluctuations disappear as the bubbling regime experiences a transition from the M2 mode to the M1 mode (Fig.~\ref{COFLOW2019_fig:Effective}$b \, IV$). It is interesting to point out that the critical pressure amplitude required to achieve mode M1 is considerably larger than that corresponding to mode M2.

All the results presented hereafter were obtained for effective forcing processes at the critical pressure amplitude, since they represent the energetically optimal values of interest for possible applications.

%--------------------------------------------------------------------------------%
%---------------------------------RESULTS----------------------------------------%
%--------------------------------------------------------------------------------%
\section{Results for the effective forcing process}\label{COFLOW2019_section:Results}

\subsection{Description of the forced bubbling process under modes M1 and M2}\label{COFLOW2019_subsection:Res_Bubbling}

The periodic bubble generation naturally established under the unperturbed flow conditions is driven by the inertia of the outer liquid stream~\citep{RodriguezARFM2015}. Indeed, the velocity at which the growing bubble interface is transported during the bubbling cycle is imposed by the co-flowing liquid~\citep{SevillaJFM2005}. Moreover, the characteristics of the air feeding system used in this study, in the absence of the speaker actuation, induce a bubble formation process with an air flow rate from the feeding chamber to the forming bubble that varies with time. Under these particular feeding conditions, the flow resistance, i.e. the pressure drop along the injection needle, plays an essential role during the initial stages of the bubble formation~\citep{OguzJFM93,Gordillo_2007}, being the latest ones characterized by a decrease of the air pressure inside the bubble~\citep{SevillaJFM2005}. The main aspects of the unforced bubbling cycle will be briefly described, and the reader is referred to~\citet{Gordillo_2007} for further details about the dynamics of bubble formation under constant pressure feeding conditions. The natural bubble formation process involves two different stages, namely the \emph{expansion} and \emph{collapse} stages. Once a bubble detaches from the intact ligament, a large pressure pulse inside the ligament induces radially outward velocities to the surrounding liquid, starting the expansion stage of a new bubble. As the bubble rapidly grows, its pressure decreases with time decelerating the liquid surrounding the bubble. In fact, at the end of the expansion stage, the bubble pressure drops below that of the liquid, inducing an inward acceleration of the liquid toward the axis and causing the formation of a neck at the interface. At this moment, the collapse stage begins, in which the liquid accelerates inwards in the neck region, and eventually results in the bubble pinch-off. The very last instants of the collapse, with part of the feeding air flowing through the neck, has been observed to be governed by the Bernoulli suction effect~\citep{SevillaJFM2005,Gordillo_2005,Dollet_2008,Bergmann_2009,Gekle_2010}.
\begin{figure}[t]
	\centering
    \includegraphics[width=1\textwidth]{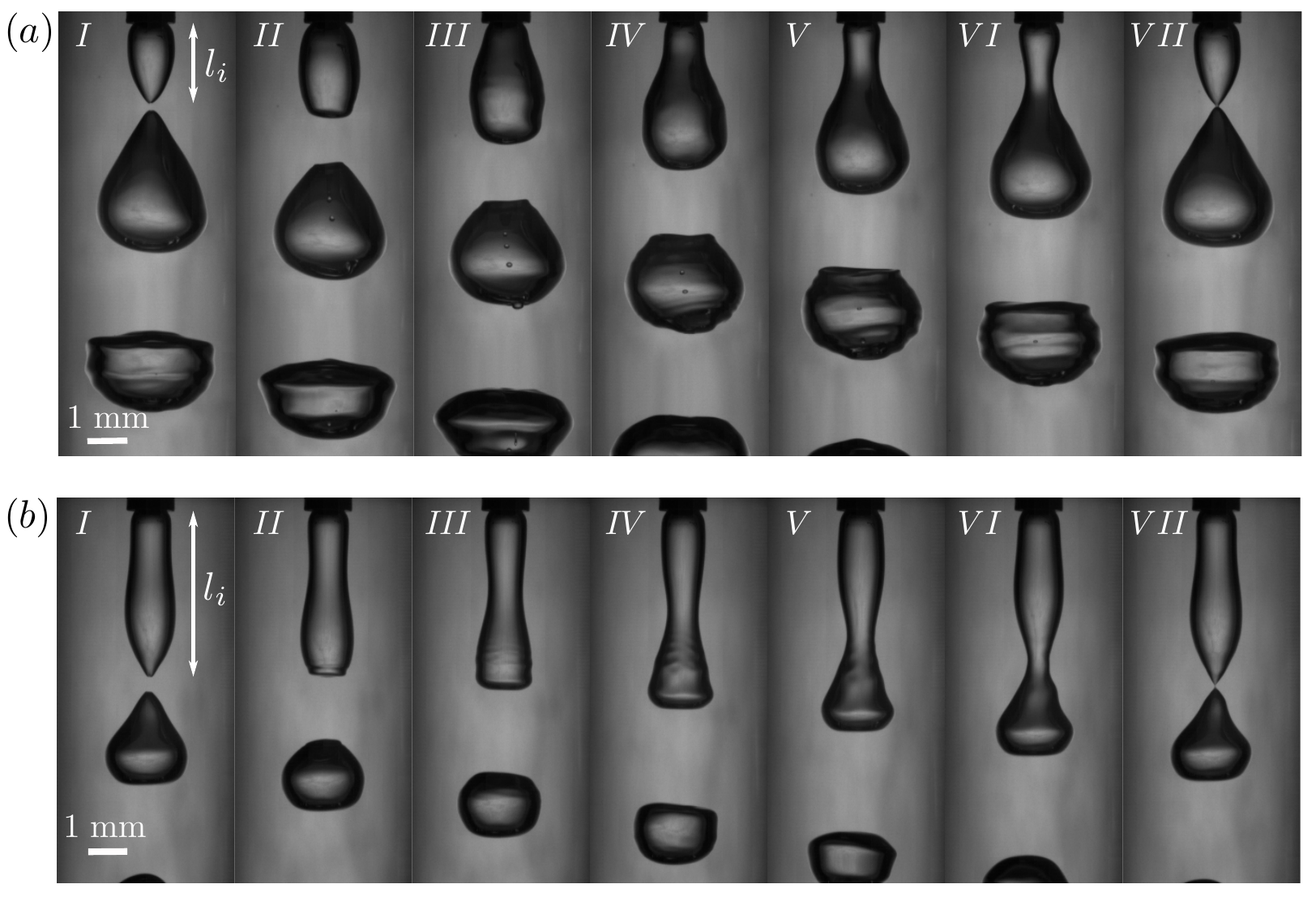}
    \caption{Sequence of experimental images showing the temporal evolution of the growing bubble during an effective forcing cycle. The shown cases include the same flow conditions as in Fig.~\ref{COFLOW2019_fig:Effective}($a,c$), respectively, forced at $f_{\!f} = 300$ Hz under the corresponding critical pressure amplitude. The obtained forced bubbling regimes exhibit two different breakup modes, ($a$) mode M1 and ($b$) mode M2, respectively. The time interval between the snapshots is 0.56 ms in both cases.}
    \label{COFLOW2019_fig:Bubbling}
\end{figure}

When an effective forcing is imposed, the pressure inside the feeding chamber varies during the bubbling cycle, as shown in Fig.~\ref{COFLOW2019_fig:Measures}, inducing a modulation of the air flow rate towards the bubble, $Q_a(t)$, that modifies the mechanisms acting in the natural case, and imposes the desired bubbling frequency, $f_{\!f}$. Figure~\ref{COFLOW2019_fig:Bubbling} shows the temporal evolution of the growing bubble interface for a complete bubbling cycle under the effect of the critical pressure amplitude for two different flow conditions, namely $u_a = 3.80$ m$/$s and $u_w = 1.36$ m$/$s in Fig.~\ref{COFLOW2019_fig:Bubbling}($a$), and $u_a = u_w = 1.36$ m$/$s in Fig.~\ref{COFLOW2019_fig:Bubbling}($b$), both forced at $f_{\!f} = 300$ Hz. The figure shows that the bubbling process in Fig.~\ref{COFLOW2019_fig:Bubbling}($a$) corresponds to the breakup mode M1, while that in Fig.~\ref{COFLOW2019_fig:Bubbling}($b$) exhibits the M2 mode. As mentioned in the previous section, from the temporal evolution of the bubble interface in both modes it is deduced that the mechanisms leading to the formation of the neck which eventually collapses, are indeed different for both modes.

The bubble formation process within the breakup mode M1 (Fig.~\ref{COFLOW2019_fig:Bubbling}$a$) clearly resembles that previously described for the natural bubbling regime. In fact, it exhibits an initial expansion stage during which the intact ligament rapidly inflates (Fig.~\ref{COFLOW2019_fig:Bubbling}$a \, I\!I$) due to the high bubble pressure. It is important to notice that, during this initial stage, the growing air cavity is not convected downstream by the outer liquid, remaining attached to the needle tip, as occurs during the formation of bubbles in a quiescent liquid pool~\citep{OguzJFM93}. However, as the bubble pressure decreases, the surrounding liquid is radially decelerated. Eventually, the air-water pressure difference induces an inward acceleration of the liquid, reducing locally the outward radial velocity of the bubble interface and giving rise to the collapse stage (Fig.~\ref{COFLOW2019_fig:Bubbling}$a \, I\!I\!I$). At this moment, the interface at the injector exit becomes parallel to the needle, and the growing air cavity begins to move downstream, generating an incipient neck between the new growing air ligament and the forming bubble (Fig.~\ref{COFLOW2019_fig:Bubbling}$a \, IV$). During this collapse stage, the neck travels downstream approximately at the water velocity and, at the same time, it accelerates towards the symmetry axis (Fig.~\ref{COFLOW2019_fig:Bubbling}$a \, V\!,V\!I$). At the end of the cycle (Fig.~\ref{COFLOW2019_fig:Bubbling}$a \, V\!I\!I$), the neck collapses, detaching a bubble from the tip of the intact ligament, whose final length is $l_i$ (Fig.~\ref{COFLOW2019_fig:Bubbling}$a \, I$). The effective nature of the forced bubbling implies that the time at which the pinch-off occurs is imposed by the forcing frequency, $1/f_b = 1/f_{\!f} < 1/f_n$. Taking into account that the intact ligament moves downwards at the water velocity only after the end of the expansion stage, the intact length in the forced case, $l_{\!f} = l_{i,f}$, is shorter than the induced perturbation wavelength, $l_{\!f} < \lambda_{\!f}$, as shall be confirmed in the next subsection. A similar behaviour is observed in the natural cases, in which the intact length, $l_n = l_{i,n}$, is also smaller than the corresponding naturally induced perturbation wavelength, $l_n < \lambda_n = u_w/f_n$. Therefore, it can be stated that the forced bubbling process within mode M1 is essentially an enhanced counterpart of the natural case, being both stages assisted by a pressure modulation that result in a faster bubbling process.

The dynamics of the M2 breakup mode (Fig.~\ref{COFLOW2019_fig:Bubbling}$b$) is different from that described both for the natural case, and for the M1 breakup mode. The main feature that can be observed in the time evolution of the growing interface, concerns the time taken by the neck to form and to grow until it collapses. In fact, when a bubble is detached (Fig.~\ref{COFLOW2019_fig:Bubbling}$b \, I$), the interface of the intact ligament is already perturbed, showing a deflection between the injector and the tip of the ligament, which corresponds to the forming neck. Similarly to the other situations, such neck travels downstream at the water velocity while it grows toward the axis (Fig.~\ref{COFLOW2019_fig:Bubbling}$b \, I\!I$-$IV$). However, in mode M2 the intact length is larger than the induced perturbation wavelength, $l_{\!f} > \lambda_{\!f}$. It is worth highlighting that, in mode M2, the initial expansion stage is substituted  by an interfacial perturbation induced by the forced inflation of the ligament. This perturbation leads to the formation of the neck (Fig.~\ref{COFLOW2019_fig:Bubbling}$b \, I\!I$) and a subsequent local inward acceleration until it pinches off (Fig.~\ref{COFLOW2019_fig:Bubbling}$b \, I\!I\!I$-$V\!I$). This implies that, contrary to what happens in mode M1 and the natural cases, the overpressure leading to the interface expansion in mode M2 is not related to the pinch-off of the previous bubble, but it is induced by the pressure perturbation created by the forcing process. Therefore, although the bubbles are emitted at the selected frequency, they take more than a forcing wavelength to grow up and detach. Thus, in mode M2 there is more than one perturbation wavelength simultaneously present in the air ligament. In addition, the necessary conditions to generate bubbles within mode M2 typically involve lower air flow rates than those required within mode M1 for equal forcing frequencies and under similar co-flow velocities. These particular conditions lead to smaller values of the required critical pressure amplitudes, as can be checked in Fig.~\ref{COFLOW2019_fig:Effective}($b,d$).

\subsection{Intact length and bubble volume}\label{COFLOW2019_subsection:Res_Volume}

\begin{figure}
	\centering
    \includegraphics[width=0.75\textwidth]{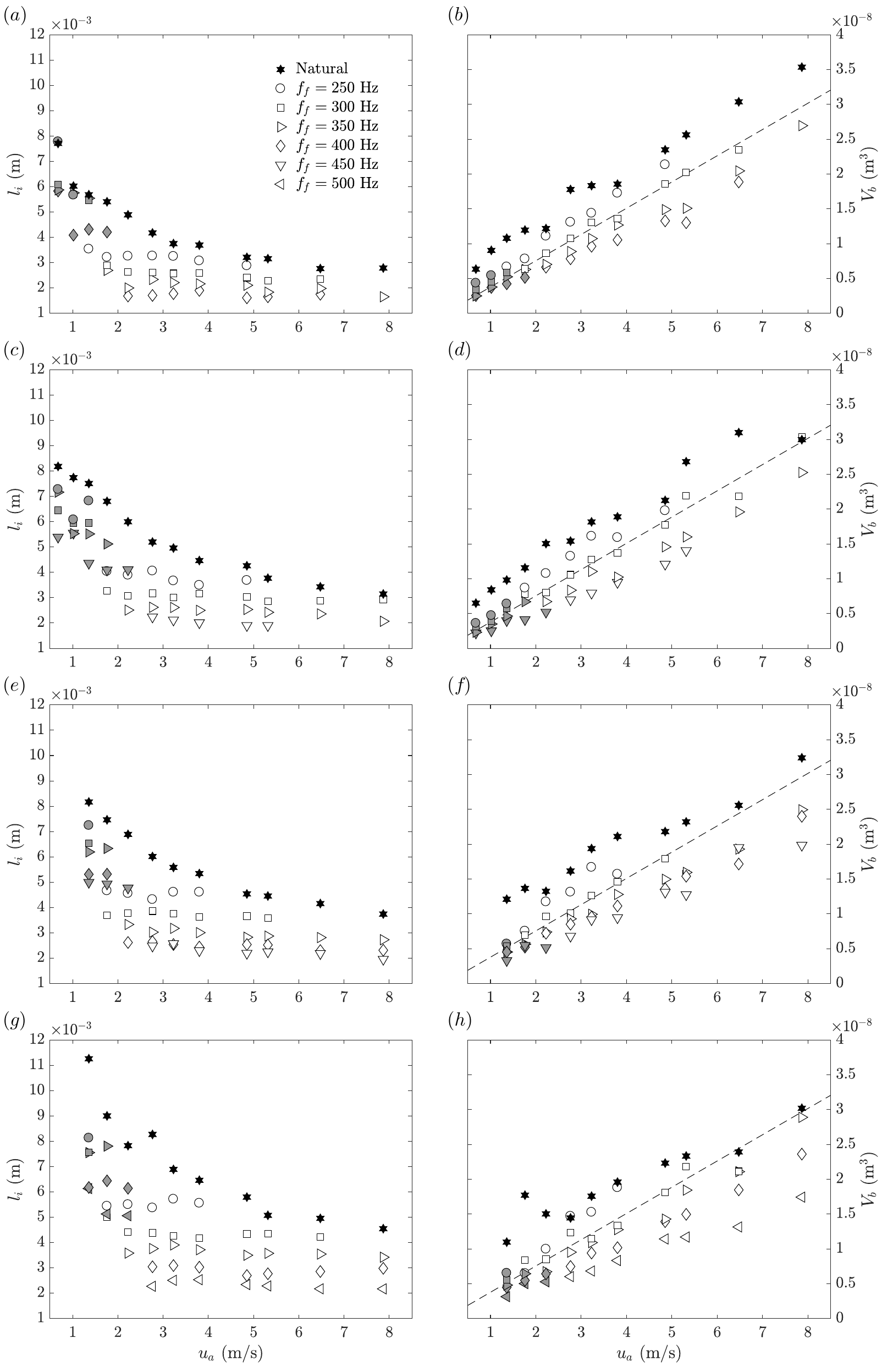}
    \caption{($a,c,e,g$) Experimental intact length and ($b,d,f,h$) bubble volume obtained for forced bubbling regimes under the critical pressure amplitude, as a function of the mean air velocity, for different forcing frequencies and ($a,b$) $u_w = 1.36$ m/s, ($c,d$) $u_w = 1.53$ m/s, ($e,f$) $u_w = 1.70$ m/s, ($g,h$) $u_w = 1.86$ m/s. For clarity, not all the forcing frequencies have been represented for each water velocity. Hollow and filled symbols correspond to the M1 and M2 breakup modes, respectively. Black stars represent the natural bubbling cases. The dashed line is the function $V_b=\pi r_o^2u_a/f_{\!f}$, being thus a straight line with a slope $\pi r_o^2/f_{\!f}$, with $f_{\!f} = 300$ Hz.}
    \label{COFLOW2019_fig:LiVb}
\end{figure}

It has been previously stated that the intact ligament length, $l_i$, represents the distance traveled by the neck from the needle tip until the position at which it pinches-off. Since for the natural and mode M1 cases the attached ligament length is approximately constant during the expansion stage, the time elapsed between both positions corresponds to the duration of the collapse stage. However, in the breakup mode M2 the neck already begins to form during the previous bubbling cycle, leading to a larger travel time, and thus to an increased intact length. Therefore, $l_i$ depends on the bubbling frequency, $f_b$, and on the axial velocity of the interface, which is close to the water velocity, $u_w$. These dependencies can be observed in Fig.~\ref{COFLOW2019_fig:LiVb}, where the intact length in natural and forced bubbling events are represented as a function of the mean air velocity for different water velocities and forcing frequencies.

Let us first focus on the natural bubbling regimes ($\bigvarstar$). It can be observed that, for constant values of the water velocity, the intact length, denoted as $l_n$ for the unperturbed case, decreases as the air velocity increases. In these cases, the bubble pinch-off takes place closer to the injector exit due to the increase of the natural bubbling frequency, as already reported by~\citep{SevillaJFM2005} for constant flow rate feeding conditions and~\citep{Gordillo_2007} for constant injection pressure conditions. On the other hand, an increase of the intact length with the water velocity is also observed (see, in addition, snapshots $I$ in Fig.~\ref{COFLOW2019_fig:Images}). 

Regarding the results for the M1 forced bubbling regime (hollow symbols), it is worth remembering that each unperturbed bubbling regime is forced at frequencies higher than the natural one, $f_{\!f}>f_n$, resulting in smaller intact ligaments, $l_{\!f}<l_{n}$. In addition, for a given value of the water velocity, a slight decrease of $l_{\!f}$ is observed with the air velocity for a constant imposed frequency. Moreover, since the mechanisms driving the forced bubble detachment within mode M1 have been described as an accelerated version of the natural case, the pinch-off occurs closer to the injector, leading to shorter intact ligaments (see also Fig.~\ref{COFLOW2019_fig:Images}$a$). Finally, the effect of the water velocity on the intact length is simply associated with the axial velocity of the neck, since the bubbling frequency is prescribed by the forcing process (see also snapshots $I\!I\!, I\!I\!I$ and $IV$ in Fig.~\ref{COFLOW2019_fig:Images}).

A different behaviour is observed in the intact lengths of the breakup mode M2, although trends similar to those observed in mode M1 are found. Generally, the intact length increases with the water velocity (see, also, snapshots $V\!I$ in Fig.~\ref{COFLOW2019_fig:Images}$b,c$), and decreases with the bubbling frequency (see, in addition, snapshots $V$ and $V\!I$ in Fig.~\ref{COFLOW2019_fig:Images}$c$), being the influence of the air velocity almost negligible. However, as explained before, the M2 intact lengths are substantially larger than those obtained within mode M1 for the same forcing frequencies (see, in addition, snapshots $I$ in Fig.~\ref{COFLOW2019_fig:Bubbling}), since the neck is already formed in the previous forcing cycle. It is important to notice that the forced intact lengths, even for mode M2, are always smaller than those corresponding to the unforced cases.

The dependence of the bubble volume, $V_b$, on the air velocity under unforced conditions is displayed in Fig.~\ref{COFLOW2019_fig:LiVb}($b,d,f,h$) with the symbol $\bigvarstar$ for the four values of the water velocity. As expected, an increase of $V_b$ with the air velocity, $u_a$, is observed. Indeed, the continuity equation and the periodicity condition imply that $V_b = \pi r_o^2 u_a / f_b$, and the bubbling frequency increases slower than the air velocity at a constant water velocity~\citep{SevillaJFM2005}. Taking into account the small range of water velocities explored in the present work, a dependence of $V_b$ with $u_w$ is barely observed. In addition, the volume of the generated bubbles under both forced regimes at different frequencies, is plotted in the same figure (hollow symbols for M1 and filled ones for M2). Since the bubble generation process is synchronized with the air flow rate modulation, a linear dependence of the bubble volume with the air velocity is obtained for constant values of the water velocity. This linear dependence is confirmed with the dashed lines, which represent the function $V_b = \pi r_o^2 u_a / f_{\!f}$ for $f_{\!f} = 300$ Hz. The good agreement between the experimental data measured at $f_{\!f} = 300$ Hz (hollow squares) and the linear function, confirms that $V_b = Q_c / f_{\!f}$, being $Q_c$ the air flow rate constantly injected into the feeding chamber. Moreover, since the bubbling frequency is imposed by the forcing process, the bubble volume does not depend on the water velocity.

\begin{figure}[t]
	\centering
    \includegraphics[width=1.0\textwidth]{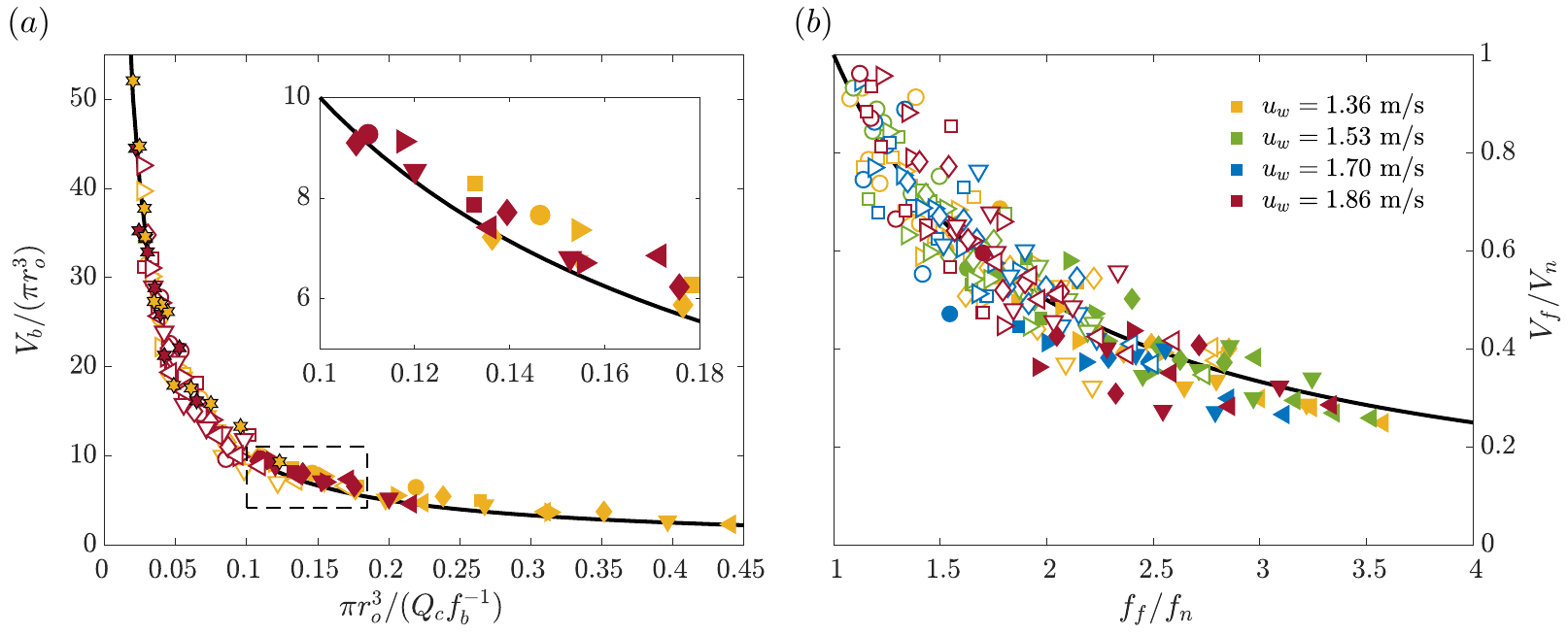}
    \caption{($a$) Dimensionless bubble volume as a function of the dimensionless frequency. The stars represent the results for the natural cases, while the rest of the symbols indicate the different forcing frequencies, as indicated in Fig.~\ref{COFLOW2019_fig:LiVb}. The solid line corresponds to $V_b= Q_c f_b^{-1}$. For clarity, only the results for the lowest and the largest values of the water velocities, $u_w = 1.36$ (yellow) and $1.86$ m$/$s (red), are shown. The inset is a zoomed representation of the region highlighted by the dashed frame, where only data from mode M2 have been plotted, to indicate the values at which mode M2 is triggered at each water velocity. ($b$) Forced-to-natural bubble volume ratio as a function of the forcing frequency ratio for all the forced cases. The solid line represents the function $V_{\!f}/V_n=f_n/f_{\!f}$. Forced bubbling events corresponding to mode M1 are represented by hollow symbols, while those associated to mode M2 are plotted with colored ones.}
    \label{COFLOW2019_fig:AdimVb}
\end{figure}

In order to fully confirm the previous statements for all the forcing frequencies, Fig.~\ref{COFLOW2019_fig:AdimVb}($a$) shows the experimentally obtained values of the bubble volume, made dimensionless with $\pi r_o^3$, as a function of the control parameters, $Q_c$ and $f_b$. Since the results for the natural cases (colored star symbols) have been included as well, the bubbling frequency $f_b$ is either $f_{b}=f_n$ in the unforced cases, or $f_{b}=f_{\!f}$, in the forced ones. In this figure, only the results for the lowest and the largest values of the experimental water velocities, $u_w = 1.36$ and $1.86$ m$/$s respectively, are plotted for clarity. It can be observed that the dimensionless bubble volume decreases with the dimensionless frequency, $\pi r_o^3 f_b/Q_c$, with all the results collapsing onto the curve $V_b=Q_c/f_b$, represented by a solid line. An important feature which can be deduced from Fig.~\ref{COFLOW2019_fig:AdimVb}($a$) is that the forced bubbling events corresponding to mode M2 (filled symbols) lead to much smaller bubble volumes than those corresponding to mode M1 (hollow symbols), since M2 typically involves flow conditions associated with lower air velocities. Indeed, mode M2 allows the production of nearly monodisperse bubbles that are even smaller than the smallest ones achieved in the natural regime, in which case the minimum air velocity is limited by the transition to the jetting regime~\citep{Sevilla_2005_PoF}. This fact, together with the small values of the critical pressure amplitude required for the effective bubbling process (see Sections~\ref{COFLOW2019_subsection:Exp_Criteria} and~\ref{COFLOW2019_subsection:Res_Bubbling}), make mode M2 potentially interesting for practical applications.

As happens to the intact length, by effectively forcing the bubbling process with frequencies higher than the natural one, the volume of the generated bubbles decreases (see Fig.~\ref{COFLOW2019_fig:LiVb}$b,d,f,h$), at the same time that the size monodispersity is improved. The latter conclusion is summarized in Fig.~\ref{COFLOW2019_fig:AdimVb}($b$), in terms of the forced-to-natural volume ratio, $V_{\!f} / V_n$, represented as a function of the forced-to-natural frequency ratio, $f_{\!f} / f_n$. Here, $V_{\!f} = V_{b,f}$ and $V_n = V_{b,n}$ represent the bubble volume of the forced and natural cases, respectively. It can be observed that the achieved volume reduction is approximately inversely proportional to the imposed frequency ratio, confirming the previous results. In addition, depending on the flow conditions and the forcing frequency, the bubble volumes can be reduced up to 80\% of their natural values, with the maximum size reduction accomplished in the forced M2 regimes, as stated above.

\subsection{Necessary conditions to achieve the M2 forced bubbling regime}\label{COFLOW2019_subsection:Res_Transition}

\begin{figure}[t]
	\centering
    \includegraphics[width=1.0\textwidth]{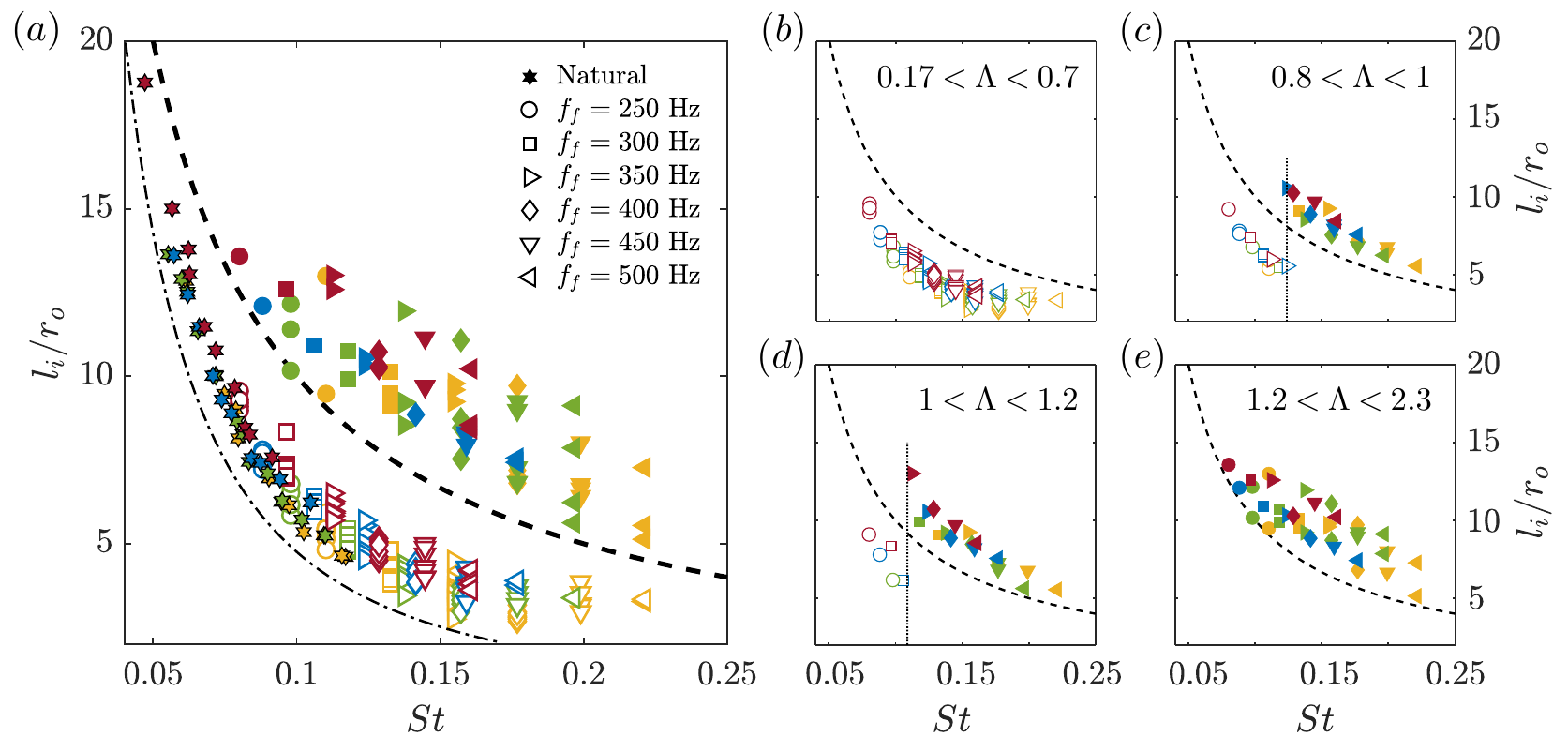}
    \caption{Dimensionless intact length as a function of the dimensionless bubbling frequency, $St = f_b r_o / u_w$. Different colors indicate the experimental water velocities, as indicated in Fig.~\ref{COFLOW2019_fig:AdimVb}. Results for the natural bubbling regimes are represented by stars, while breakup modes M1 and M2 are plotted as hollow and solid symbols, respectively. The dashed line represents the dimensionless perturbation wavelength, $l_i/r_o = St^{-1}$. ($a$) Results for all the experimental data, including the natural cases. The thin dashed-dotted line is a function $l_i/r_o \sim St^{-8/5}$ (see Eq.~\ref{COFLOW2019_eq:LN} in Sect.~\ref{COFLOW2019_section:Char_Length}). ($b$-$e$) Results for different ranges of $\Lambda$. The vertical dotted lines point out the approximate values of $St$ at which the transition from mode M1 to M2 takes place in each panel.}
    \label{COFLOW2019_fig:AdimLi}
\end{figure}
In order to qualitatively explore the conditions under which the forcing process leads to bubble production within the breakup mode M2, it can be firstly stated from the experimental images shown in Fig.~\ref{COFLOW2019_fig:Images}, that they depend on the air and water velocities, as well as on the forcing frequency. Moreover, the frequency at which mode M2 is triggered varies with the unperturbed flow conditions. Therefore, it proves convenient to represent its dependence in compact form using dimensionless variables, namely $\Lambda$ and $St$. In addition, it has been shown that mode M1 is characterized by a forced intact length shorter than the corresponding perturbation wavelength, $l_{\!f} < \lambda_{\!f}$, while $l_{\!f} > \lambda_{\!f}$ in the M2 mode. In this sense, Fig.~\ref{COFLOW2019_fig:AdimLi}($a$) shows the dimensionless intact length, $l_i/r_o$, as a function of the dimensionless bubbling frequency, defined as the bubbling Strouhal number, $St = f_b r_o / u_w$, for all the experiments performed in this work, including the natural and the forced cases. It is worth mentioning again that, under the critical pressure amplitude, the bubbling frequency is the forcing one, $f_{b}=f_{\!f}$ and, consequently, $St=St_{\!f}$. The intact length for the unperturbed cases monotonically decreases with $St$ with the same trend for all the analyzed values of the water velocity, represented with different colors. Moreover, as stated before, the mechanisms leading to the formation of the neck in the breakup mode M1 are the same as in the natural bubbling regime. Thus, the forced intact lengths obtained for the M1 mode (hollow symbols) follow the same behaviour as those for the natural cases (stars). In fact, similar values of $St$ give rise to similar intact lengths in both cases, confirming that the M1 regime behaves as an artificially assisted natural case. Nevertheless, some data dispersion is observed in Fig.~\ref{COFLOW2019_fig:AdimLi}($a$) due to the effect of the water velocity. An additional data dispersion is also apparent at each forcing frequency, showing the slight decrease of the forced intact length with $u_a$, previously noticed in Fig.~\ref{COFLOW2019_fig:LiVb}($a,c,e,g$). A clear distinction of the two different breakup modes can be seen in this figure. Indeed, the intact lengths obtained from the natural bubbling regimes and from the break mode M1, lie below the dimensionless perturbation wavelength, $u_w/(f_b r_o)  = St^{-1}$, represented with a dashed line. The difference reflects the fraction of the bubbling cycle spent on the initial expansion stage. On the other hand, the intact lengths of mode M2 bubbling regimes are longer than the perturbation wavelength, since the neck is formed during the previous bubbling cycle. A larger data dispersion is present the M2 events, due to the slenderness of the ligament and the larger receptivity to noise disturbances. Therefore, as anticipated above, it is concluded that the breakup mode M2 takes place under flow conditions which lead to intact ligaments longer than the induced perturbation wavelength, $l_{\!f} = \alpha \, \lambda_{\!f}$, being $\alpha > 1$.

It is well known that the periodic bubble formation, naturally established for gas-liquid co-flowing jets, is associated with an absolute instability~\citep{Monkewitz_1988,Sevilla_2005_PoF}, in which the characteristic time associated with the local growth of interfacial disturbances is smaller than the liquid convective time, $r_o/u_w$. However, as the water-to-air velocity ratio, $\Lambda$, increases, the convective time decreases, giving rise to bubble pinch-off at increasing distances from the injector. Therefore, larger values of $\Lambda$ lead to longer intact ligaments, promoting the appearance of the breakup mode M2 under short enough forced perturbation wavelengths. This can be confirmed in Figs.~\ref{COFLOW2019_fig:AdimLi}($b$)-($e$), where the results have been classified in different ranges of $\Lambda$. Notice that for the lowest range of $\Lambda$, shown in Fig.~\ref{COFLOW2019_fig:AdimLi}($b$), the bubbling events correspond to breakup mode M1 under the whole range of values of $St$ analyzed (see, in addition, Fig.~\ref{COFLOW2019_fig:Images}$a$). In this case, where the air velocity is considerable faster than the liquid velocity, the air stream is slowed down by the co-flowing water stream when it exits the injector, and expands radially to accommodate the incoming gas flow rate. Thus, an expansion stage takes place before forming a neck near the needle exit, initiating then the collapse stage. As $\Lambda$ increases (Fig.~\ref{COFLOW2019_fig:AdimLi}$c,d$), the forced regimes respond either to mode M1 or mode M2, depending on the value of the induced $St$ (see, in addition, Figs.~\ref{COFLOW2019_fig:Images}$b,c$). In fact, there is a critical value of the Strouhal number, $St_c$, indicated with vertical dotted lines in Figs.~\ref{COFLOW2019_fig:AdimLi}($c$)-($d$), at which a transition from mode M1 to mode M2 occurs. It can be observed that $St_c$ decreases as $\Lambda$ increases, confirming the previously mentioned relationship between the naturally established wavelength, associated to the water-to-air velocity ratio, and the forced perturbation wavelength, caused by the induced $St$. Finally, for large enough values of $\Lambda$ (Fig.~\ref{COFLOW2019_fig:AdimLi}$e$), the forced bubbling process always takes place in mode M2 under the whole range of $St$ analyzed here. In this case, the velocity of the water stream is larger than that of the air and, once it exits the injector, the gas stream flows parallel to the surrounding liquid, being the system convectively unstable~\citep{Sevilla_2005_PoF}. Thus, the gas jet breaks further downstream than in the case of mode M1, assisted by the high amplitude perturbations introduced by the forcing system.

As mentioned above, Figs.~\ref{COFLOW2019_fig:AdimLi}($b$)-($e$), indicate that the critical frequency (or $St_c$) for the transition between both forced breakup modes varies with $\Lambda$. In fact, plotting all the experimental results in the $St-\Lambda$ plane (Fig.~\ref{COFLOW2019_fig:STvsL}), it can be seen that modes M1 and M2 occupy different regions in the parametric plane, being the transition from mode M1 to M2 in the shaded area. Consequently, in order to clearly determine the transition between both modes, a detailed analysis of the results in Fig.~\ref{COFLOW2019_fig:AdimVb}($a$) reveals the particular conditions at which the mode M2 is triggered. In particular, noting that the dimensionless gas volume injected to the feeding chamber, $\pi r_o^3 / (Q_c f_{\!f}^{-1})$, can be rewritten as $St \Lambda$, the transition between modes is observed to occur for the range $0.1 \lesssim St_c \Lambda \lesssim 0.13$ (see inset in Fig.~\ref{COFLOW2019_fig:AdimVb}$a$). Therefore, the critical Strouhal number can be bounded by $0.1\Lambda^{-1} \lesssim St_c \lesssim 0.13 \Lambda^{-1}$, as shown by the shaded area in Fig.~\ref{COFLOW2019_fig:STvsL}. More precisely, it can be observed that this critical value evolves from the upper to the lower bound as $\Lambda$ increases. In fact, the inset in Fig.~\ref{COFLOW2019_fig:AdimVb}($a$) reveals that $St_c$ not only depends on $\Lambda$, but its exact value also depends on the the water velocity, or similarly, on $W\!e$, as shall be discussed in Sect.~\ref{COFLOW2019_section:Char_Length}.

\section{Characterization of the air intact length and transition between breakup modes}\label{COFLOW2019_section:Char_Length}

\begin{figure}[t]
	\centering
    \includegraphics[width=0.7\textwidth]{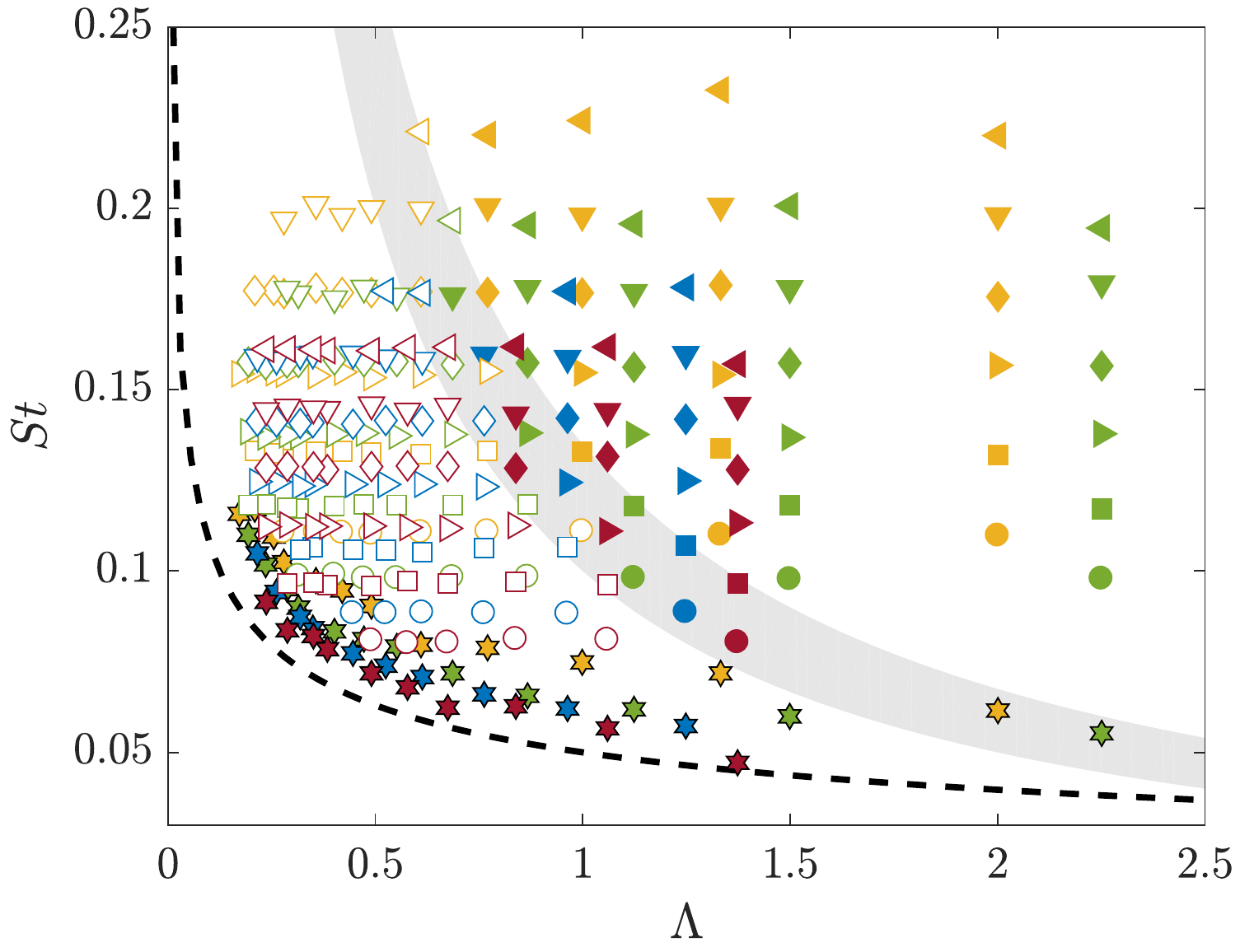}
    \caption{Experimental bubbling events obtained in the present work in the $St - \Lambda$ plane. Results for the natural bubbling regimes are represented as colored stars and the rest of the symbols denote the different forcing frequencies, as indicated in Fig.~\ref{COFLOW2019_fig:AdimVb}. Hollow symbols indicate the results corresponding to breakup mode M1 and the solid ones to mode M2, respectively. Different colors indicate the experimental water velocities, as indicated in Fig.~\ref{COFLOW2019_fig:AdimVb}. The shaded area represents the region bounded by $0.1\Lambda^{-1} \leq St \leq 0.13 \Lambda^{-1}$ where the transition between the breakup modes takes place. The dashed line is the function $St \sim \Lambda^{-1/3}$ (see Sect.~\ref{COFLOW2019_section:Char_Length}).}
    \label{COFLOW2019_fig:STvsL}
\end{figure}

\begin{figure}[t]
	\centering
    \includegraphics[width=1.0\textwidth]{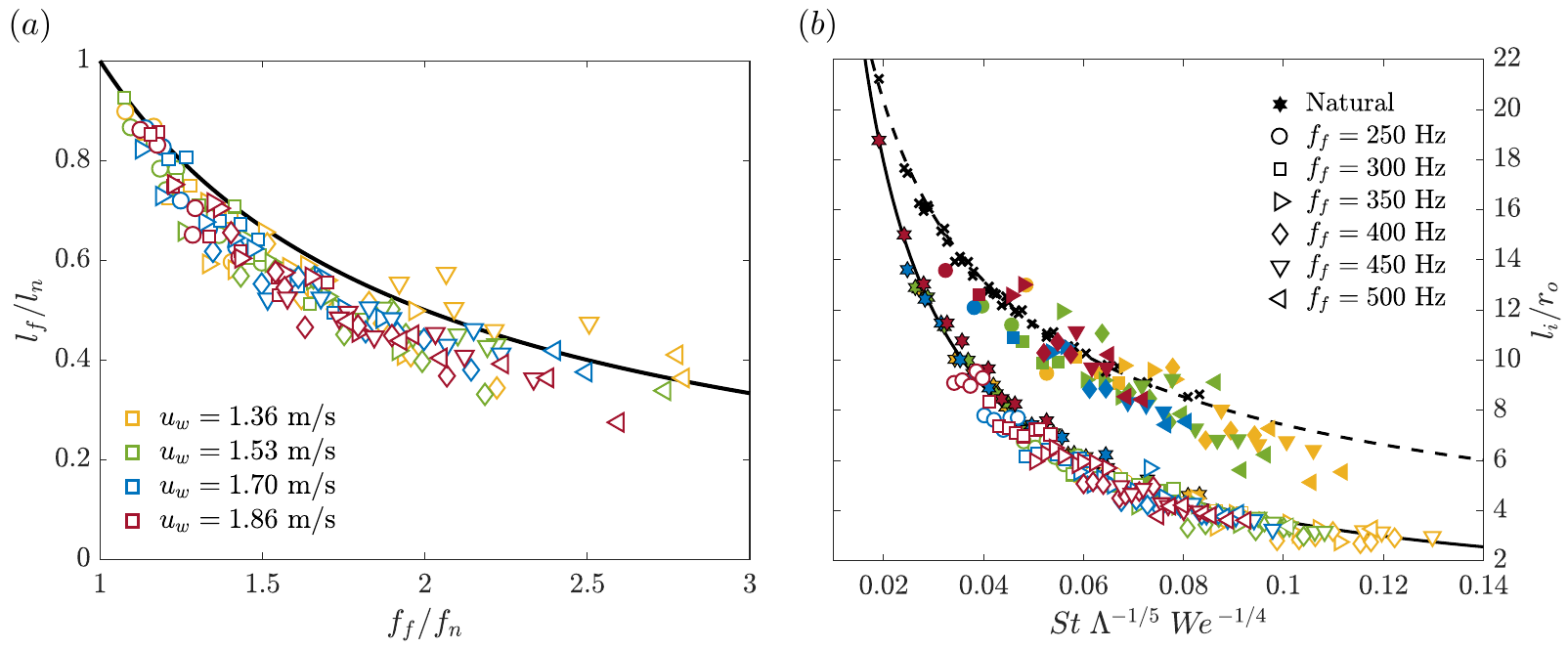}
    \caption{($a$) Forced-to-natural intact length ratio as a function of the forced-to-natural frequency ratio for all the forced bubbling processes in the M1 breakup regime. The solid line represents the function $l_{\!f}/l_n = f_n/f_{\!f}$. ($b$) Dependence of the dimensionless intact length, $l_i/r_o$, on the dimensionless control parameters, namely the bubbling frequency, $St$, and the flow conditions, specified by $\Lambda$ and $W\!e$, for all the experiments. The continuous line represents the dimensionless intact length given by Eq.~\eqref{COFLOW2019_eq:Li_Scale2}. The black crosses denote the unperturbed dimensionless wavelength, $\lambda_n/r_o=(u_w/f_n)/ r_o$. The dashed line represents the dimensionless wavelength of a natural case, characterized by $\Lambda$ and $W\!e$, whose dimensionless frequency is $St_n \sim \Lambda^{-1/3} We^{-1/3}$, given by Eq.~\eqref{COFLOW2019_eq:StN2}.}
    \label{COFLOW2019_fig:LiCharct}
\end{figure}

The experimental results presented in the previous sections clearly demonstrate that the bubble formation dynamics in the breakup mode M1 resembles that taking place in the unforced bubbling regime. Indeed, the bubbling time, imposed by $f_{\!f}$ in the forced case, is composed by the expansion and the collapse stages observed in the natural bubbling process. Each stage establishes the conditions that periodically lead to the other one, being their duration dependent on the unperturbed flow conditions and, in the forced case, on both the forcing frequency and amplitude. These dependencies imply that the effective pressure modulation assists both stages, accelerating the process to generate bubbles at the imposed frequency. Moreover, it has been observed that the relative time spent on each stage can be determined by the intact length. Therefore, similarly to the experimental forced-to-natural bubble volume ratio, both the natural and the forced M1 bubbling processes can be compared in terms of the ratio between the forced and natural intact lengths, $l_{\!f}/l_n$, for each imposed frequency ratio, $f_{\!f}/f_n$. Thus, Fig.~\ref{COFLOW2019_fig:LiCharct}($a$) shows that the experimental data nearly follows the function $l_{\!f}/l_n = (f_{\!f}/f_n)^{-1}$, represented by a solid line, suggesting that the mechanisms driving the forced bubbling process are equivalent to those acting in the natural case. Therefore, the intact lengths obtained in the forced cases can be characterized as a function of the control parameters through a comparison with their natural counterparts, making use of the relationship extracted from Fig.~\ref{COFLOW2019_fig:LiCharct}($a$).

To that aim, let us first identify the dependence of the intact length obtained in the natural bubbling cases with the unperturbed flow conditions. As suggested by Fig.~\ref{COFLOW2019_fig:STvsL}, for all values of the water velocity, the natural dimensionless bubbling frequency, $St_n = f_n r_o / u_w$, depends on $\Lambda$. Indeed, it has been found that, for each value of the water velocity (colored stars in Fig.~\ref{COFLOW2019_fig:STvsL}), $St_n$ scales as $\Lambda^{-1/3}$, as corroborated by the dashed line in Fig.~\ref{COFLOW2019_fig:STvsL}. The scatter observed among the data series associated with different water velocities may well reflect a slight dependence with the Weber number~\citep{Gordillo_2007}. Indeed, the surface tension forces contribute to accelerate the neck collapse. Therefore, the dimensionless bubbling frequency obtained under the unperturbed flow conditions can be appropriately approximated by
\begin{equation}
St_n \sim \Lambda^{-1/3} \mathscr{F},
\label{COFLOW2019_eq:StN}
\end{equation}
where $\mathscr{F} = \mathscr{F}(W\!e)$ is an \textit{a priori} unknown function of the Weber number. In a similar way, the analysis of the dimensionless natural intact length, $l_n/r_o$, reveals a similar behaviour for each water velocity (colored stars in Fig.~\ref{COFLOW2019_fig:AdimLi}$a$). In fact, it has been found that 
\begin{equation}
l_n/r_o \sim St_n^{-8/5} \mathscr{G},
\label{COFLOW2019_eq:LN}
\end{equation}
where $\mathscr{G} = \mathscr{G}(W\!e)$ is an additional unknown function of the Weber number (see Fig.~\ref{COFLOW2019_fig:AdimLi}$a$). A closer inspection of the experimental results leads to $\mathscr{G} \simeq W\!e^{1/20}$, showing a very weak dependence of the intact length on the Weber number. In addition, the fact that the absolute value of the exponent of $St_n$ is larger than unity, reflects the dependence of the relative time spent on both bubbling stages with the dimensionless bubbling frequency. 
For a given flow condition, defined by the values of $\Lambda$ and $W\!e$, as well as a dimensionless forcing frequency, $St_{\!f} = f_{\!f} r_o / u_w$, the dimensionless forced intact length, $l_{\!f}/r_o$, can be characterized by including Eqs.~(\ref{COFLOW2019_eq:StN}) and~(\ref{COFLOW2019_eq:LN}) into the expression $l_{\!f}/l_n = f_n/f_{\!f}$, extracted from Fig.~\ref{COFLOW2019_fig:LiCharct}($a$), providing
\begin{equation}
l_{\!f}/r_o \sim St_{\!f}^{-1} \Lambda^{1/5} \mathscr{H}.
\label{COFLOW2019_eq:Li_Scale}
\end{equation} 
In Eq.~\eqref{COFLOW2019_eq:Li_Scale}, $\mathscr{H} = \mathscr{F}^{-3/5} \mathscr{G}$ is a new unknown function which includes the dependence of $l_n/r_o$ and of $St_n$ on $W\!e$. The experimental results indicate that $\mathscr{H} \simeq W\!e^{1/4}$, and consequently $\mathscr{F} \simeq W\!e^{-1/3}$. Taking into account the dependencies of $\mathscr{F}$ and $\mathscr{H}$ on $W\!e$, Eqs.~\eqref{COFLOW2019_eq:StN}-\eqref{COFLOW2019_eq:Li_Scale} can be expressed as,
\begin{eqnarray}
& &St_n \, W\!e^{1/3}\sim \Lambda^{-1/3},\label{COFLOW2019_eq:StN2}\\
& &l_n/r_o \sim St_n^{-8/5} W\!e^{1/20} ,\label{COFLOW2019_eq:LN2}\\
& &l_{\!f} /r_o \sim St_{\!f}^{-1} \Lambda^{1/5} W\!e^{1/4}.\label{COFLOW2019_eq:Li_Scale2}
\end{eqnarray}

\begin{figure}[t]
	\centering
    \includegraphics[width=0.7\textwidth]{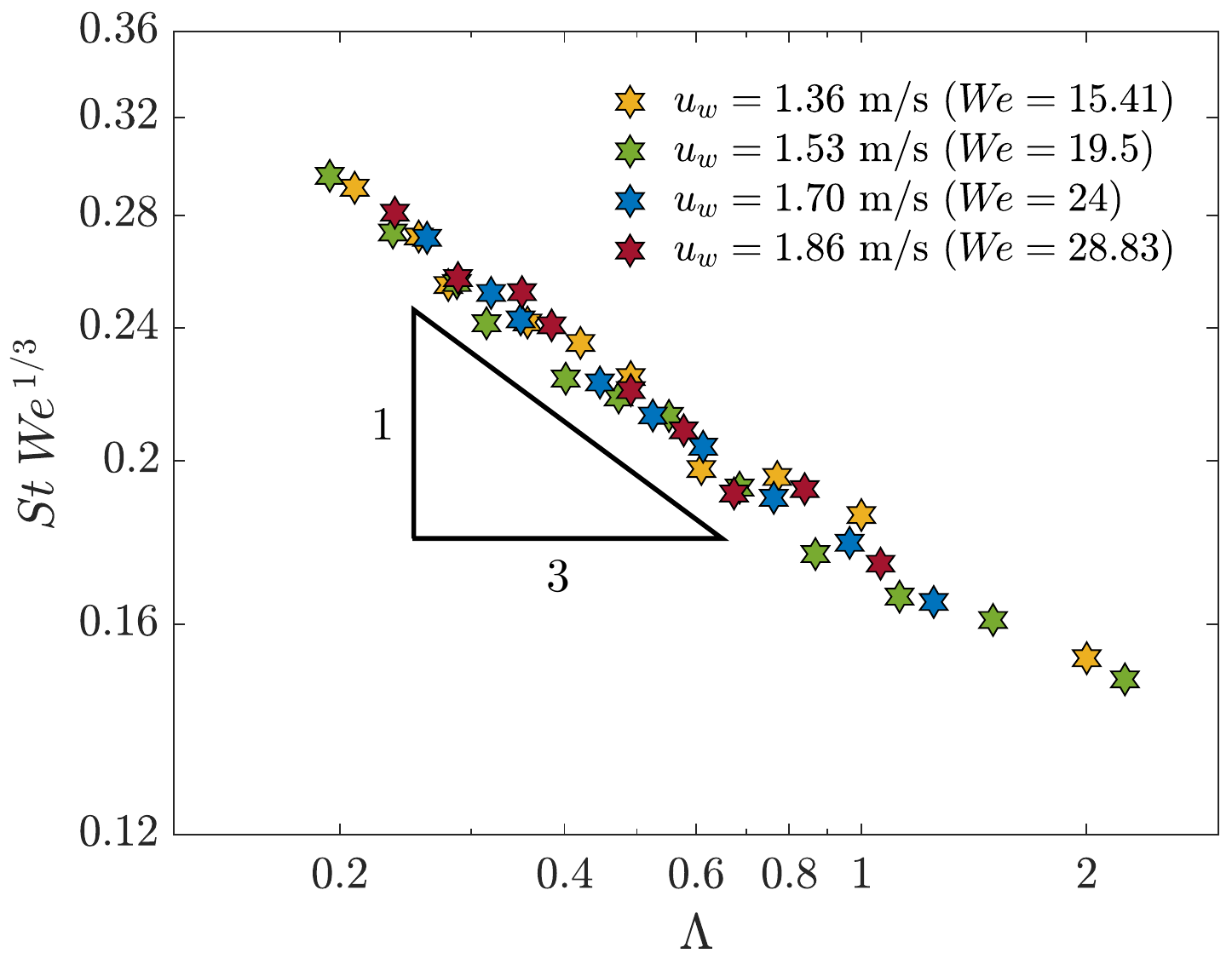}
    \caption{Dimensionless bubbling frequency corrected by $W\!e^{1/3}$ for the natural cases as a function of $\Lambda$ represented in logarithmic scale and grouped by the different experimental water velocities.}
    \label{COFLOW2019_fig:STvsL_B}
\end{figure}
To check the validity of these expressions, Fig.~\ref{COFLOW2019_fig:STvsL_B} shows, in a doubly logarithmic plot, the dimensionless natural bubbling frequency, corrected with $\mathscr{F}^{-1}(W\!e)=W\!e^{1/3}$, as a function of $\Lambda$ for all the water velocities, depicted with different colors. It can be observed that all the experimental results collapse onto a single curve with a slope of $-1/3$, as determined by Eq.~\eqref{COFLOW2019_eq:StN2}. Moreover, this result confirms the validity of the function $\mathscr{F}(W\!e)$ in Eq.~(\ref{COFLOW2019_eq:StN}).

\begin{figure}[t]
	\centering
    \includegraphics[width=0.7\textwidth]{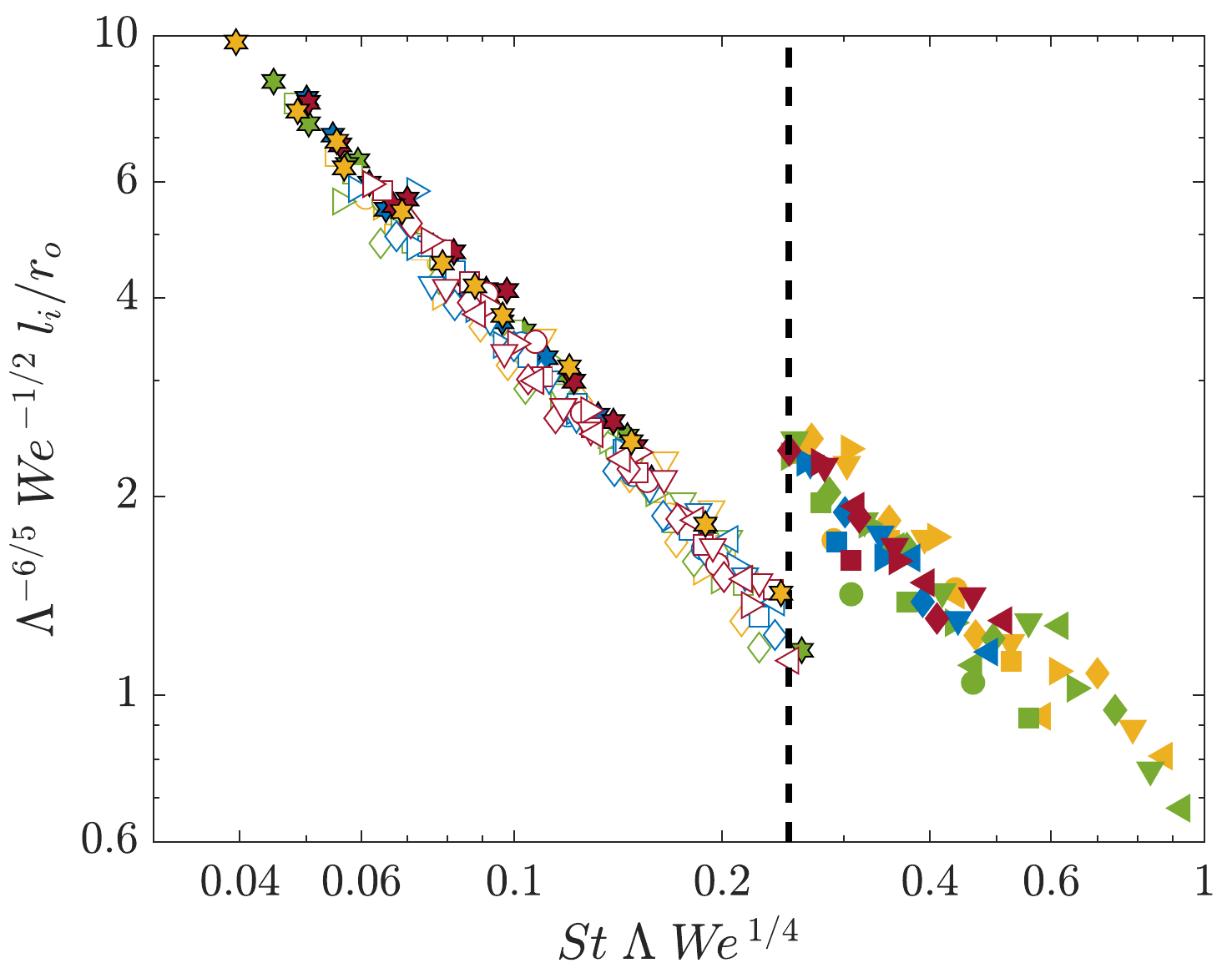}
    \caption{Doubly logarithmic plot of the compensated dimensionless intact length as a function of $St \Lambda W\!e^{1/4}$. The sharp transition between the breakup modes M1 (open symbols) and M2 (solid symbols) is marked with a vertical dashed line. The results of the natural cases are included as colored stars, while the rest of the symbols indicate the different forcing frequencies, as indicated in Fig.~\ref{COFLOW2019_fig:AdimVb}. The different experimental values of $W\!e$ are plotted with different colors, as indicated in Fig.~\ref{COFLOW2019_fig:STvsL_B}.}
    \label{COFLOW2019_fig:Transition}
\end{figure}
Figure~\ref{COFLOW2019_fig:LiCharct}($b$) shows the dimensionless intact length versus the product $St \Lambda^{-1/5} W\!e^{-1/4}$, for all the natural cases (colored stars) and the forced ones (open and solid symbols, corresponding to the breakup modes M1 and M2, respectively). It can be observed that the unforced results, as well as the forced ones associated to the breakup mode M1, collapse onto a single curve, given by Eq.~\eqref{COFLOW2019_eq:Li_Scale2}. It is worth pointing out that previous works devoted to forced breakup processes in planar co-flowing air-water streams~\citep{Ruiz-Rus_2017}, as well as co-flowing liquid-liquid systems~\citep{ZhuPA_2016}, have reported a decrease of the intact length as the inverse of the dimensionless forcing frequency. However, the present work provides an additional correction to include the dependence of the intact length on the flow conditions, i.e.~$\Lambda$ and $W\!e$, which is related to the expansion stage of the bubbling cycle. In addition, the dimensionless intact length obtained with Eq.~(\ref{COFLOW2019_eq:Li_Scale2}) is plotted as a solid line in Fig.~\ref{COFLOW2019_fig:LiCharct}($b$), corroborating the agreement with the experimental results. Notice that this correlation can be interpreted as the intact length that would be obtained in the natural bubbling regime established for the flow conditions described by $\Lambda$ and $W\!e$, with a dimensionless bubbling frequency $St_n$ given by Eq.~(\ref{COFLOW2019_eq:StN2}), as determined by Eq.~(\ref{COFLOW2019_eq:LN2}). The good agreement obtained for the forced cases within the M1 regime confirms the fact that the mechanisms driving the forced bubble formation process in the breakup mode M1 are the same as those governing the unforced case. Moreover, note that the dimensionless wavelength of the natural cases (black crosses in Fig.~\ref{COFLOW2019_fig:LiCharct}$b$), given by $u_w/(f_n r_o)$, are always larger than the intact length, indicating that a non-negligible fraction of the bubbling time is spent on the expansion stage. Similarly, the dimensionless wavelength of the equivalent natural cases, obtained as $St_n^{-1} \sim (\Lambda W\!e)^{1/3}$ according to Eq.~\eqref{COFLOW2019_eq:StN2}, is depicted by a dashed line in Fig.~\ref{COFLOW2019_fig:LiCharct}($b$). As previously mentioned, this equivalent dimensionless wavelength would coincide with the intact length in the absence of the expansion stage. Consequently, the experimental results obtained for the forced cases corresponding the breakup mode M2 (solid symbols) lie very close to the equivalent dimensionless wavelength (dashed line).

Therefore, as evidenced by Fig.~\ref{COFLOW2019_fig:LiCharct}, the proposed scale for the dimensionless intact length in Eq.~\eqref{COFLOW2019_eq:Li_Scale2}, namely $l_{\!f}/r_o \sim St_{\!f}^{-1} \Lambda^{1/5} W\!e^{1/4}$, accurately describes the parametric dependence for all cases, since it includes the effect of both the bubbling frequency, $St=St_{\!f}$, as well as the flow conditions, $\Lambda$ and $W\!e$. In addition, this scale can be used to determine the critical conditions associated with the transition between the M1 and M2 breakup modes. To that end, the results shown in Fig.~\ref{COFLOW2019_fig:AdimVb}($a$) and Fig.~\ref{COFLOW2019_fig:LiCharct}($b$) are reassembled in the doubly logarithmic plot shown in Fig.~\ref{COFLOW2019_fig:Transition}. Here, the dimensionless intact length $l_i/r_o$, compensated with the product $\Lambda^{-6/5} W\!e^{-1/2}$, is represented as a function of $St \Lambda W\!e^{1/4}$, as suggested by correlation~\eqref{COFLOW2019_eq:Li_Scale2}. The compensated intact length is seen to monotonically decrease with $St \Lambda W\!e^{1/4}$, until a sharp transition takes place from the breakup mode M1 (open symbols) to the breakup mode M2 (solid symbols) at ($St \Lambda W\!e^{1/4})_c = 0.25$, as indicated by the vertical dashed line in Fig.~\ref{COFLOW2019_fig:Transition}, providing $St_c=0.25 \,W\!e^{-1/4} \Lambda^{-1} $. Beyond this critical value, the compensated intact length suddenly increases, and then monotonically decreases again, showing the emission of smaller bubbles from an air ligament of larger length than in the M1 regime. Considering the range of Weber numbers covered in this study $15 < W\!e < 33$, we get $0.1 < 0.25 \,W\!e^{-1/4}< 0.13$ as already commented in Sect.~\ref{COFLOW2019_subsection:Res_Transition} and reported in Fig.~\ref{COFLOW2019_fig:AdimLi}. Moreover, this result also describes the shaded area in Fig.~\ref{COFLOW2019_fig:STvsL}, where the transition is observed to take place for a $St_c(\Lambda)$ in the range $0.1\Lambda^{-1} < St_c <0.13\Lambda^{-1}$, depending the exact value on $W\!e$. In addition, it can be noticed that the M2 mode allows the production of bubbles in ranges that cannot be reached within the unforced bubbling regime (colored stars). Thus, the production of bubbles within the M2 breakup mode overcomes the minimum bubble volume limitation imposed by the injector geometry associated with the natural bubbling flow, at the same time that it avoids the lack of monodispersity typically associated with convectively unstable regimes.

\section{Conclusions}\label{COFLOW2019_section:Conclusions}

This work describes a new method to actively control the bubbling process in axisymmetric air-water co-flowing jets. With this technique, the bubble generation frequency, as well as the volume of the obtained monodisperse bubbles, can be independently controlled. Consequently, bubbles substantially smaller than those formed in the unperturbed case are generated, reducing the influence of the injector geometry. To that aim, a forcing system based on a loudspeaker, that induces a periodic pressure modulation in the air stream, has been incorporated into the air feeding line of a cylindrical gas-liquid injector. 

Its performance during the bubbling process has been experimentally analyzed from synchronized measurements of the growing bubble interface extracted from high-speed visualizations and of the pressure inside the feeding chamber. Large enough amplitudes of the pressure modulation have been found to control the bubble formation process, generating monodisperse bubbles of volume $V_b = Q_c/f_{\!f}$, at the forcing frequency $f_{\!f}$. The minimum pressure amplitude required for the forcing process to be effective was determined performing systematic experiments varying the control parameters, i.e. $W\!e$, $\Lambda$ and $St_f$. When effective, the forced bubbling process exhibits two different regimes, characterized by two different breakup modes, namely modes M1 and M2, which also depend on the unperturbed flow conditions. More precisely, the breakup mode M1 has been identified as a bubbling regime similar to the unperturbed case in which the mechanisms leading to the bubble formation are accelerated by the induced air pressure modulation, generating bubbles at the forcing frequency. On the other hand, the breakup mode M2 resembles a forced jetting regime in which an induced interface perturbation convectively grows, periodically forcing the detachment of monodisperse bubbles from the tip of a slender air ligament. It has been observed that mode M2 allows the production of very uniform bubbles under flow conditions which are not achievable for the corresponding unperturbed case, thus obtaining small monodisperse bubbles whose volume overcomes the limitation imposed by the geometry of the injector. Bubble volume reductions up to $80\%$ of the corresponding volume obtained for the same unperturbed flow conditions, are shown to be achieved in the range of parameters analyzed in this work.

In addition, the established forced bubbling regimes have been characterized by means of the intact length, which is the distance at which the bubble pinch-off takes place. This length properly describes the duration of both stages involved in the bubble formation process, assuming that the induced interface perturbation travels downstream at the outer co-flowing liquid stream velocity. A scaling law, given by $l_i/r_o \sim St_f^{-1} \Lambda^{1/5} W\!e^{1/4}$ has been proposed to accurately describe the dependence of the dimensionless intact length, $l_i/r_o$, on the governing parameters of the bubbling process. Finally, a representation of this correlation for all the experimental conditions covered in the present study has allowed us to determine the particular conditions under which the breakup mode M2 is triggered. Indeed $(St_f \Lambda W\!e)_c=0.25$ has been found to be the condition for the transition from mode M1 to mode M2.
%---------------------------------CRediT authorship contribution statement-----------------------%
\section*{CRediT authorship contribution statement}
{\bf J. Ruiz-Rus:} Conceptualization, Methodology, Software, Investigation, Visualization, Writing - original draft. {\bf R. Bola\~nos-Jim\'enez:} Conceptualization, Supervision, Visualization, Writing - review \& editing, Funding acquisition. {\bf A. Sevilla}: Conceptualization, Supervision, Writing - review \& editing, Funding acquisition. {\bf C. Mart\'{\i}nez-Baz\'an:} Conceptualization, Supervision, Project administration, Writing - review \& editing, Funding acquisition.
%----------------------------------Declaration of Competing Interest --------------------%
\section*{Declaration of Competing Interest}
The authors declare that they have no known competing financial interests or personal relationships that could have appeared to influence the work reported in this paper.

%--------------------------------ACKNOWLEDGEMENTS-------------------------------%
\section*{Acknowledgments}
This work has been supported by the Spanish MINECO and European Funds under projects DPI2017-88201-C3-2-R and DPI2017-88201-C3-3-R. JRR wants to acknowledge the Spanish MINECO for the financial support provided by the Fellowship BES-2015-071329.

%------------------------------------BIBLIOGRAPHY--------------------------------%
%\bibliographystyle{elsarticle-harv}
%\bibliography{Tesis_biblio}

\begin{thebibliography}{45}
\expandafter\ifx\csname natexlab\endcsname\relax\def\natexlab#1{#1}\fi
\expandafter\ifx\csname url\endcsname\relax
  \def\url#1{\texttt{#1}}\fi
\expandafter\ifx\csname urlprefix\endcsname\relax\def\urlprefix{URL }\fi

\bibitem[{Abe and Sanada(2015)}]{Abe_ChemEngSci_2015}
Abe, K., Sanada, T., 2015. The mechanism of bubble generation using a slit
  elastic tube and an acoustic pressure wave in the gas phase. Chem. Eng. Sci.
  128, 28--35.

\bibitem[{{\AA}mand et~al.(2013){\AA}mand, Olsson, and
  Carlsson}]{Amand_Review_2013}
{\AA}mand, L., Olsson, G., Carlsson, B., 2013. Aeration control--a review.
  Water Sci. Technol. 67~(11), 2374--2398.

\bibitem[{Basaran(2002)}]{Basaran2002}
Basaran, O.~A., 2002. Small-scale free surface flows with breakup: drop
  formation and emerging applications. AIChE J. 48, 1842--1848.

\bibitem[{Bergmann et~al.(2009)Bergmann, Andersen, Van~der Meer, and
  Bohr}]{Bergmann_2009}
Bergmann, R., Andersen, A., Van~der Meer, D., Bohr, T., 2009. Bubble pinch-off
  in a rotating flow. Phys. Rev. Lett. 102~(20), 204501.

\bibitem[{Bhunia et~al.(1998)Bhunia, Pais, Komotani, and Kim}]{Bhunia_1998}
Bhunia, A., Pais, S., Komotani, Y., Kim, I., 1998. Bubble formation in a coflow
  configuration in normal and reduced gravity. AIChE J. 44, 1499--1509.

\bibitem[{Bola{\~n}os-Jim{\'e}nez et~al.(2016)Bola{\~n}os-Jim{\'e}nez, Sevilla,
  and Mart{\'\i}nez-Baz{\'a}n}]{Bolanos_IJMF2016}
Bola{\~n}os-Jim{\'e}nez, R., Sevilla, A., Mart{\'\i}nez-Baz{\'a}n, C., 2016.
  Modeling of the bubbling process in a planar co-flow configuration. Int. J.
  Multiphase Flow 82, 86--92.

\bibitem[{Canny(1986)}]{Canny_1986}
Canny, J., 1986. A computational approach to edge detection. IEEE Trans.
  Pattern Anal. Mach. Intell.~(6), 679--698.

\bibitem[{Chuang and Goldschmidt(1970)}]{Chuang1970}
Chuang, S.~C., Goldschmidt, V.~W., 1970. Bubble formation due to a submerged
  capillary tube in quiescent and coflowing streams. J. Basic Engng. 92~(4),
  705--711.

\bibitem[{Corchero et~al.(2006)Corchero, Medina, and Higuera}]{Corchero_2006}
Corchero, G., Medina, A., Higuera, F.~J., 2006. Effect of wetting conditions
  and flow rate on bubble formation at orifices submerged in water. Colloids
  Surf. A 290~(1-3), 41--49.

\bibitem[{Di~Bari and Robinson(2013)}]{DiBari_2013}
Di~Bari, S., Robinson, A.~J., 2013. Adiabatic bubble growth in uniform {DC}
  electric fields. Exp. Therm Fluid Sci. 44, 114--123.

\bibitem[{Dollet et~al.(2008)Dollet, Van~Hoeve, Raven, Marmottant, and
  Versluis}]{Dollet_2008}
Dollet, B., Van~Hoeve, W., Raven, J.-P., Marmottant, P., Versluis, M., 2008.
  Role of the channel geometry on the bubble pinch-off in flow-focusing
  devices. Phys. Rev. Lett. 100~(3), 034504.

\bibitem[{Ferrara et~al.(2007)Ferrara, Pollard, and Borden}]{Ferrara_2007}
Ferrara, K., Pollard, R., Borden, M., 2007. Ultrasound microbubble contrast
  agents: fundamentals and application to gene and drug delivery. Annu. Rev.
  Biomed. Eng. 9, 415--447.

\bibitem[{Fujikawa et~al.(2003)Fujikawa, Zhang, Hayama, and
  Peng}]{Fujikawa_2003}
Fujikawa, S., Zhang, R., Hayama, S., Peng, G., 2003. The control of
  micro-air-bubble generation by a rotational porous plate. Int. J. Multiphase
  Flow 29~(8), 1221--1236.

\bibitem[{Gekle et~al.(2010)Gekle, Peters, Gordillo, van~der Meer, and
  Lohse}]{Gekle_2010}
Gekle, S., Peters, I.~R., Gordillo, J.~M., van~der Meer, D., Lohse, D., 2010.
  Supersonic air flow due to solid-liquid impact. Phys. Rev. Lett. 104~(2),
  024501.

\bibitem[{Gonz{\'a}lez and Woods(2002)}]{Gonzalez_2002}
Gonz{\'a}lez, R.~C., Woods, R.~E., 2002. Digital image processing.
  Prentice-Hall, {USA}.

\bibitem[{Gordillo et~al.(2007)Gordillo, Sevilla, and
  Mart{\'\i}nez-Baz{\'a}n}]{Gordillo_2007}
Gordillo, J.~M., Sevilla, A., Mart{\'\i}nez-Baz{\'a}n, C., 2007. Bubbling in a
  co-flow at high {R}eynolds numbers. Phys. Fluids 19~(7), 077102.

\bibitem[{Gordillo et~al.(2005)Gordillo, Sevilla,
  Rodr{\'\i}guez-Rodr{\'\i}guez, and Mart{\'\i}nez-Baz{\'a}n}]{Gordillo_2005}
Gordillo, J.~M., Sevilla, A., Rodr{\'\i}guez-Rodr{\'\i}guez, J.,
  Mart{\'\i}nez-Baz{\'a}n, C., 2005. Axisymmetric bubble pinch-off at high
  {R}eynolds numbers. Phys. Rev. Lett. 95~(19), 194501.

\bibitem[{Grinis and Monin(1999)}]{Grinis_1999}
Grinis, L., Monin, Y., 1999. Influence of vibrations on gas bubble formation in
  liquids. Chem. Eng. Technol. 22~(5), 439--442.

\bibitem[{Guti{\'e}rrez-Montes et~al.(2013)Guti{\'e}rrez-Montes,
  Bola{\~n}os-Jim{\'e}nez, Sevilla, and
  Mart{\'\i}nez-Baz{\'a}n}]{Gutierrez-MontesIJMF2013}
Guti{\'e}rrez-Montes, C., Bola{\~n}os-Jim{\'e}nez, R., Sevilla, A.,
  Mart{\'\i}nez-Baz{\'a}n, C., 2013. Experimental and numerical study of the
  periodic bubbling regime in planar co-flowing air--water sheets. Int. J.
  Multiphase Flow 50, 106--119.

\bibitem[{Guti{\'e}rrez-Montes et~al.(2014)Guti{\'e}rrez-Montes,
  Bola{\~n}os-Jim{\'e}nez, Sevilla, and
  Mart{\'\i}nez-Baz{\'a}n}]{Gutierrez-MontesIJMF2014}
Guti{\'e}rrez-Montes, C., Bola{\~n}os-Jim{\'e}nez, R., Sevilla, A.,
  Mart{\'\i}nez-Baz{\'a}n, C., 2014. Bubble formation in a planar
  water--air--water jet: {E}ffects of the nozzle geometry and the injection
  conditions. Int. J. Multiphase Flow 65, 38--50.

\bibitem[{Hijano et~al.(2015)Hijano, Loscertales, Ib{\'a}{\~n}ez, and
  Higuera}]{Hijano_2015}
Hijano, A.~J., Loscertales, I.~G., Ib{\'a}{\~n}ez, S.~E., Higuera, F.~J., 2015.
  Periodic emission of droplets from an oscillating electrified meniscus of a
  low-viscosity, highly conductive liquid. Phys. Rev. E 91~(1), 013011.

\bibitem[{Jim{\'e}nez-Gonz{\'a}lez and
  Huera-Huarte(2017)}]{Jimenez-Gonzalez_2017}
Jim{\'e}nez-Gonz{\'a}lez, J.~I., Huera-Huarte, F.~J., 2017. Experimental
  sensitivity of vortex-induced vibrations to localized wake perturbations. J.
  Fluids Struct. 74, 53--63.

\bibitem[{Lee(1974)}]{Lee_JResDev_1974}
Lee, H.~C., 1974. Drop formation in a liquid jet. IBM J. Res. Dev. 18~(4),
  364--369.

\bibitem[{Leib and Goldstein(1986{\natexlab{a}})}]{LeibyGoldsteinAC}
Leib, S., Goldstein, M., 1986{\natexlab{a}}. Convective and absolute
  instability of a viscous liquid jet. Phys. Fluids 29~(4), 952--954.

\bibitem[{Leib and Goldstein(1986{\natexlab{b}})}]{LeibyGoldstein}
Leib, S., Goldstein, M., 1986{\natexlab{b}}. The generation of capillary
  instabilities on a liquid jet. J. Fluid Mech. 168, 479--500.

\bibitem[{Makuta et~al.(2013)Makuta, Suzuki, and
  Nakao}]{Makuta_Ultrasonic_2013}
Makuta, T., Suzuki, R., Nakao, T., 2013. Generation of microbubbles from hollow
  cylindrical ultrasonic horn. Ultrasonics 53~(1), 196--202.

\bibitem[{Monkewitz and Sohn(1988)}]{Monkewitz_1988}
Monkewitz, P.~A., Sohn, K., 1988. Absolute instability in hot jets. AIAA J.
  26~(8), 911--916.

\bibitem[{Najafi et~al.(2008)Najafi, Xu, and Masliyah}]{Najafi_2008}
Najafi, A.~S., Xu, Z., Masliyah, J., 2008. Single micro-bubble generation by
  pressure pulse technique. Chem. Eng. Sci. 63~(7), 1779--1787.

\bibitem[{O{\~g}uz and Prosperetti(1993)}]{OguzJFM93}
O{\~g}uz, H.~N., Prosperetti, A., 1993. Dynamics of bubble growth and
  detachment from a needle. J. Fluid Mech. 257, 111--145.

\bibitem[{Ostmann and Schwarze(2018)}]{Ostmann_2018}
Ostmann, S., Schwarze, R., 2018. A compact device for the deterministic
  generation of medium-sized bubbles. Rev. Sci. Instrum. 89~(12), 125108.

\bibitem[{Rodr{\'\i}guez-Rodr{\'\i}guez
  et~al.(2015)Rodr{\'\i}guez-Rodr{\'\i}guez, Sevilla, Mart{\'\i}nez-Baz{\'a}n,
  and Gordillo}]{RodriguezARFM2015}
Rodr{\'\i}guez-Rodr{\'\i}guez, J., Sevilla, A., Mart{\'\i}nez-Baz{\'a}n, C.,
  Gordillo, J.~M., 2015. Generation of microbubbles with applications to
  industry and medicine. Annu. Rev. Fluid Mech. 47, 405--429.

\bibitem[{Ruiz-Rus et~al.(2017)Ruiz-Rus, Bola{\~n}os-Jim{\'e}nez,
  Guti{\'e}rrez-Montes, Sevilla, and Mart{\'\i}nez-Baz{\'a}n}]{Ruiz-Rus_2017}
Ruiz-Rus, J., Bola{\~n}os-Jim{\'e}nez, R., Guti{\'e}rrez-Montes, C., Sevilla,
  A., Mart{\'\i}nez-Baz{\'a}n, C., 2017. Controlled formation of bubbles in a
  planar co-flow configuration. Int. J. Multiphase Flow 89, 69--80.

\bibitem[{Sanada and Abe(2013)}]{Sanada_2013}
Sanada, T., Abe, K., 2013. Generation of single bubbles of various sizes using
  a slitting elastic tube. Rev. Sci. Instrum. 84~(8), 085106.

\bibitem[{Sevilla et~al.(2005{\natexlab{a}})Sevilla, Gordillo, and
  Mart{\'\i}nez-Baz{\'a}n}]{SevillaJFM2005}
Sevilla, A., Gordillo, J.~M., Mart{\'\i}nez-Baz{\'a}n, C., 2005{\natexlab{a}}.
  Bubble formation in a coflowing air--water stream. J. Fluid Mech. 530,
  181--195.

\bibitem[{Sevilla et~al.(2005{\natexlab{b}})Sevilla, Gordillo, and
  Mart{\'\i}nez-Baz{\'a}n}]{Sevilla_2005_PoF}
Sevilla, A., Gordillo, J.~M., Mart{\'\i}nez-Baz{\'a}n, C., 2005{\natexlab{b}}.
  Transition from bubbling to jetting in a coaxial air--water jet. Phys. Fluids
  17~(1), 018105.

\bibitem[{Shirota et~al.(2008)Shirota, Sanada, Sato, and
  Watanabe}]{Shirota_PoF_2008}
Shirota, M., Sanada, T., Sato, A., Watanabe, M., 2008. Formation of a
  submillimeter bubble from an orifice using pulsed acoustic pressure waves in
  gas phase. Phys. Fluids 20~(4), 043301.

\bibitem[{Song and Springer(1996)}]{Song_1996}
Song, B., Springer, J., 1996. Determination of interfacial tension from the
  profile of a pendant drop using computer-aided image processing: 2.
  {E}xperimental. J. Colloid Interface Sci. 184~(1), 77--91.

\bibitem[{Tsuge et~al.(1997)Tsuge, Tanaka, Terasaka, and Matsue}]{Tsuge_1997}
Tsuge, H., Tanaka, Y., Terasaka, K., Matsue, H., 1997. Bubble formation in
  flowing liquid under reduced gravity. Chem. Eng. Sci. 52~(21-22), 3671--3676.

\bibitem[{Vega et~al.(2009)Vega, Montanero, and Fern{\'a}ndez}]{Vega_2009}
Vega, E.~J., Montanero, J.~M., Fern{\'a}ndez, J., 2009. On the precision of
  optical imaging to study free surface dynamics at high frame rates. Exp.
  Fluids 47~(2), 251--261.

\bibitem[{Vejrazka et~al.(2008)Vejrazka, Fujasov{\'a}, Stanovsky, Ruzicka, and
  Draho{\v{s}}}]{Vejrazka_FDR_2008}
Vejrazka, J., Fujasov{\'a}, M., Stanovsky, P., Ruzicka, M.~C., Draho{\v{s}},
  J., 2008. Bubbling controlled by needle movement. Fluid Dyn. Res. 40~(7-8),
  521.

\bibitem[{Waghmare et~al.(2008)Waghmare, Rice, and
  Knopf}]{Waghmare_IndEngChemRes_2008}
Waghmare, Y., Rice, R., Knopf, F., 2008. Mass transfer in a viscous bubble
  column with forced oscillations. Ind. Eng. Chem. Res. 47~(15), 5386--5394.

\bibitem[{Wang et~al.(2016)Wang, Chen, Yuan, Wang, Li, Zhang, and
  Liu}]{Wang_2016}
Wang, N., Chen, X., Yuan, J., Wang, G., Li, Y., Zhang, H., Liu, Y., 2016.
  Bubble formation at a submerged orifice in high-speed horizontal oscillation.
  Metall. Mater. Trans. B 47~(6), 3362--3374.

\bibitem[{Wang et~al.(2018)Wang, Hernan, and Chen}]{Wang_2018}
Wang, S., Hernan, B.~J., Chen, C.-L., 2018. Towards enhanced bubble detachment
  within a thin liquid film by electrowetting with voltage modulation. Phys.
  Fluids 30~(6), 062102.

\bibitem[{Xu et~al.(2017)Xu, Yan, Wang, and Chen}]{Xu_2017}
Xu, H., Yan, R., Wang, S., Chen, C.-L., 2017. Bubble detachment assisted by
  electrowetting-driven interfacial wave. Phys. Fluids 29~(10), 102105.

\bibitem[{Zhu et~al.(2016)Zhu, Tang, and Wang}]{ZhuPA_2016}
Zhu, P., Tang, X., Wang, L., 2016. Droplet generation in co-flow microfluidic
  channels with vibration. Microfluid. Nanofluid. 20~(3), 47.

\end{thebibliography}
%\section*{References}

\end{document}